\documentclass[twocolumn]{aastex631}

\newcommand{\kms}{\ensuremath{\mathrm{km}\,\mathrm{s}^{-1}}}

\shorttitle{Gas dynamics and star formation in NGC 6822}
\shortauthors{Park et al.}

\graphicspath{{./}{figures/}}

\usepackage{amsmath}
\usepackage{lipsum}
\usepackage{enumitem}

\begin{document}

\title{Gas dynamics and star formation in NGC 6822}

\newcommand\blfootnote[1]{%
  \begingroup
  \renewcommand\thefootnote{}\footnote{#1}%
  \addtocounter{footnote}{-1}%
  \endgroup
}

\author[0000-0002-9809-6631]{Hye-Jin Park$^{**}$}
\blfootnote{$^{**}$Current address: Research School of Astronomy and Astrophysics, Australian National University, Cotter Road, Weston Creek, ACT 2611, Australia}
\affiliation{Department of Physics and Astronomy, Sejong University, 209 Neungdong-ro, Gwangjin-gu, Seoul 05006, Republic of Korea}

\author[0000-0002-8379-0604]{Se-Heon Oh}
\thanks{seheon.oh@sejong.ac.kr}
\affiliation{Department of Physics and Astronomy, Sejong University, 209 Neungdong-ro, Gwangjin-gu, Seoul 05006, Republic of Korea}

\author[0000-0002-6593-8820]{Jing Wang}
\affiliation{Kavli Institute for Astronomy and Astrophysics (KIAA), Peking University, Beijing 100871, China}
\affiliation{Department of Astronomy, School of Physics, Peking University, Beijing 100871, China}

\author{Yun Zheng}
\affiliation{Kavli Institute for Astronomy and Astrophysics (KIAA), Peking University, Beijing 100871, China}
\affiliation{Department of Astronomy, School of Physics, Peking University, Beijing 100871, China}

\author[0000-0003-1632-2541]{Hong-Xin Zhang}
\affiliation{Key Laboratory for Research in Galaxies and Cosmology, Department of Astronomy, University of Science and Technology of China, Hefei, Anhui 230026, China}
\affiliation{School of Astronomy and Space Science, University of Science and Technology of China, Hefei, Anhui 230026, China}

\author[0000-0001-8957-4518]{W. J. G. de Blok}
\affiliation{Netherlands Institute for Radio Astronomy (ASTRON), Oude Hoogeveensedijk 4, 7991 PD, Dwingeloo, The Netherlands}
\affiliation{Department of Astronomy, University of Cape Town, Private Bag X3, Rondebosch 7701, South Africa}
\affiliation{Kapteyn Astronomical Institute, University of Groningen, Postbus 800, 9700 AV Groningen, The Netherlands}

\correspondingauthor{Se-Heon Oh}

\begin{abstract}

We present H{\sc i} gas kinematics and star formation activities of NGC 6822, a dwarf galaxy located in the Local Group at a distance of $\sim$\,490 kpc. We perform profile decomposition of line-of-sight velocity profiles of the H{\sc i} data cube (42.4\arcsec $\times$ 12.0\arcsec\,spatial, corresponding to $\sim$\,100\,pc; 1.6\,\kms\,spectral) taken with the Australia Telescope Compact Array (ATCA). For this, we use a new tool, the so-called {\sc baygaud} which is based on Bayesian analysis techniques, allowing us to decompose a line-of-sight velocity profile into an optimal number of Gaussian components in a quantitative manner. We classify the decomposed H{\sc i} gas components of NGC 6822 into cool-bulk, warm-bulk, cool-non-bulk and warm-non-bulk motions with respect to their centroid velocities and velocity dispersions. We correlate their gas surface densities with corresponding star formation rate densities derived using both the GALEX far-ultraviolet and WISE 22\,$\mu$m data to examine the resolved Kennicutt-Schmidt (K-S) law for NGC 6822. Of the decomposed H{\sc i} gas components, the cool-bulk component is likely to better follow the linear extension of the K-S law for molecular hydrogen (H$_{2}$) at low gas surface densities where H{\sc i} is not saturated.

\end{abstract}


\section{Introduction}

\label{sec:intro}
Stars are expected to form in cold molecular hydrogen (H$_{2}$) gas clumps collapsed by gravitational instability which can be further fragmented into gaseous cores that may result in individual stars (\citealt{bonnell1998formation}; \citealt{truelove1998self}; \citealt{krumholz2005formation}; \citealt{williams2000structure}). In the standard view of star formation, cooling of the interstellar medium (ISM) is one of the most critical processes in making H$_{2}$ gaseous clumps or cores, which is closely associated with star formation efficiency. Despite significant progress in the understanding of the star formation process in galaxies (e.g., \citealt{bergin2007cold}; \citealt{di2007star}; \citealt{kennicutt2012star}), its detailed mechanism has not been fully understood yet. We still need to obtain more concrete information about the star formation timescale and its efficiency which critically depend on the physical properties of the ISM and environmental effects including gas turbulence, gas accretion process or gravitational interaction with other galaxies (\citealt{georgakakis2000cold}).

The Kennicutt-Schmidt star formation law (K-S law) (\citealt{schmidt1959rate}; \citealt{kennicutt1998global}; \citealt{kennicutt2012star}) describes the relationship between the surface density of gas ($\Sigma_{\rm GAS}$; how much gas in solar masses per $\rm pc^{2}$) and that of star formation rate ($\Sigma_{\rm SFR}$; how much stars form in solar masses per year per $\rm kpc^{2}$) using a power law given by
\begin{equation}
\label{eq:kslaw}
    \Sigma_{\rm SFR} = \Sigma_{\rm GAS}^{N}\,,
\end{equation}
\noindent where {\it N} is $\sim$\,1.40 $\pm$ 0.15 from \cite{kennicutt1998global}.

This empirical global star formation law has been found to extend over not only the gas and SFR densities in several orders of magnitude but the physical scales of star forming regions from a few kpc down to pc scales. In this regard, several studies have investigated the K-S law at different physical scales: 1) global (total $\Sigma_{\rm GAS}$ \--- $\Sigma_{\rm SFR}$ of galaxies; \citealt{kennicutt1998global}, \citealt{de2019revisiting}, \citealt{kennicutt2021revisiting}), 2) radially-averaged (radial $\Sigma_{\rm GAS}$ \--- $\Sigma_{\rm SFR}$ of a galaxy; \citealt{kennicutt1989star}, \citealt{wong2002relationship}, \citealt{heyer2004molecular}, \citealt{schuster2007complete}), and 3) sub-kpc scales ($\Sigma_{\rm GAS}$ \--- $\Sigma_{\rm SFR}$ of giant molecular gas clouds, GMCs; \citealt{kennicutt2007star}, \citealt{bigiel2008star}, \citealt{wyder2009star}, \citealt{roychowdhury2009star}, \citealt{bigiel2010extremely}, \citealt{onodera2010breakdown}, \citealt{roychowdhury2015spatially}, \citealt{kreckel2018scale}). 

Significant progress has been made towards understanding of galaxy star formation process at small scales as high-quality spatial and spectral resolution data become available from multi-wavelength (H{\sc i} 21cm to FUV) observations of nearby galaxies in the local Universe ($<$10 Mpc) (e.g., THINGS\footnote{The H{\sc i} Nearby Galaxy Survey}\---\citealt{walter2008}; LITTLE THINGS\footnote{Local Irregulars That Trace Luminosity Extremes THINGS}\---\citealt{hunter2012}; LVHIS\footnote{The Local Volume H{\sc i} Survey}\---\citealt{koribalski2018}; BIMA SONG\footnote{The BIMA Survey of Nearby Galaxies}\---\citealt{helfer2003}; HERACLES\footnote{The HERA CO Line Extragalactic Survey}\---\citealt{leroy2009}; SINGS\footnote{The Spitzer Nearby Galaxies Survey}\---\citealt{kennicutt2003}; GALEX Nearby Galaxies Survey\---\citealt{de2007galex}). Several studies including \cite{kennicutt2007star}, \cite{bigiel2008star}, \citet{leroy2008star}, \cite{bigiel2010extremely} and \cite{liu2011super} investigate the relationship between the $\Sigma_{\rm GAS}$ (which is the sum of $\Sigma_{HI}$ and $\Sigma_{H2}$, presenting H{\sc i} and H$_{2}$ gas surface densities, respectively) and $\Sigma_{\rm SFR}$ of nearby galaxies on a few hundreds of pc scales by combining the high-quality multi-wavelength data (e.g., H{\sc i} 21cm, CO, near-infrared, mid-infrared and far/near-ultraviolet). This so-called resolved K-S law has made a significant contribution to our understanding of the physical properties of the ISM and its link to star formation process in galaxies on individual GMC scales. Together, it provides critical observational constraints being used for modelling star formation in high-resolution hydrodynamical simulations of galaxies (\citealt{springel2003cosmological}; \citealt{calura2020hydrodynamic}; \citealt{sormani2020simulations}; \citealt{oh2020calibration}).

\cite{bigiel2008star} derive a tight relationship between the $\Sigma_{\rm H2}$ and $\Sigma_{\rm SFR}$ in the H$_{2}$-dominated centers of seven THINGS spiral galaxies and 11 late-type dwarf galaxies. It is described by a Schmidt-type power law with {\it N} = 1.00 $\pm \rm$ 0.10 in Eq.~(\ref{eq:kslaw}). This indicates a  constant gas depletion time ($\tau_{dep} = \Sigma_{\rm GAS} / \Sigma_{\rm SFR}$) in the inner disks of spirals. On the other hand, the relation breaks down in the outskirts of the spiral galaxies and in late-type dwarf galaxies where H{\sc i} dominates, and CO low-J emissions, the general tracer of H$_2$, are hardly detectable. In these H{\sc i}-dominated regions, significant scatter is seen in gas depletion time.

As found in previous studies, the $\rm N$ in Eq.~(\ref{eq:kslaw}) becomes larger than the one of the molecular K-S law at below $\Sigma_{\rm HI} \sim$10 $\rm M_{\odot}\,$\rm pc$^{-2}$ which $\Sigma_{\rm HI}$ saturates due to the conversion of a cold phase of H{\sc i} in the clouds to the molecular gas (\citealt{schaye2008relation}; see also Figure 15 in \citealt{bigiel2008star} and Figure 18 in \citealt{wyder2009star}). In the low gas surface density regime below this critical value, a significant scatter in the atomic hydrogen K-S law is seen. Additionally, the complex kinematics and structure of star-forming ISM in galaxies could contribute to the scatter in the K-S law. Gas outflows driven by hydrodynamical processes in a galaxy such as stellar winds and supernova explosions, and/or tidal interaction with other galaxies can give rise to turbulent and random motions in the ambient ISM, deviating from the global kinematics of the galaxy. Locally, the energy deposited into the ISM by stellar feedback is able to make bubbles and holes with sizes of sub-kpc to more than kpc scales (\citealt{van1996bubbles}; \citealt{begatakos2011fine}; \citealt{pokhrel2020catalog}). This can cause multi-phase structure of the ISM, which makes it difficult to derive the physical properties of the ISM and relate them to star formation process in galaxies (\citealt{hopkins2014galaxies}; \citealt{fierlinger2016stellar}).

From an observational point of view, the physical properties of H{\sc i} gaseous components in a galaxy can be derived by modelling their spectral line profiles. However, the shape of line profiles is often non-Gaussian with multiple kinematic components. This is usually caused by hydrodynamic and gravitational forces from star formation and gravitational interactions with other galaxies as well as gas streaming motions by bars and spiral arms. Such non-Gaussian velocity profiles are even found in low-inclined galaxies which are free from the projection effect (e.g., \citealt{koch2018kinematics}). Conventional profile analysis like the moment analysis and single Gaussian fitting method is limited in modelling the non-Gaussian line profiles (e.g., \citealt{oh2008high}). These conventional methods could either under- or overestimate the velocity dispersion and/or integrated flux density values of individual line-of-sight non-Gaussian velocity profiles, depending on their noise characteristics. This can result in the scatter in the resolved K-S law, particularly at low gas densities. In this respect, a proper analysis of non-Gaussian H{\sc i} line profiles of galaxies is essential for examining the relationship between $\Sigma_{\rm GAS}$ and $\Sigma_{\rm SFR}$ of galaxies in the low gas density regime.

\cite{de2006star} perform profile decomposition of H{\sc i} line profiles of NGC 6822 which is a nearby dwarf galaxy at a distance of $\sim$490 kpc from Australia Telescope Compact Array (ATCA) observations ($\sim$\,42.4\arcsec $\times$\,12\arcsec\,spatial; 1.6\,\kms\,spectral)  (\citealt{de2000evidence}).  \cite{de2006star} fit two Gaussian functions to individual H{\sc i} line profiles of a regridded ($\sim$\,48.0\arcsec $\times$\,48.0\arcsec) H{\sc i} data cube of NGC 6822, and classify them into kinematically cool (narrow; $v_{di\!s\!p}\!<$ 6\,\kms) and warm (broad; $v_{di\!s\!p}\!>$ 6\,\kms) components according to their velocity dispersion ($v_{di\!s\!p}$), respectively. Kinematically cool H{\sc i} gas is expected to be more associated with star formation as the atomic hydrogen should pass through it before turning H$_{2}$ (\citealt{young2003star}; \citealt{de2006star}; \citealt{ianjamasimanana2012shapes}; \citealt{oh2019robust}). By combining the H{\sc i} data with H$\alpha$ and optical ({\it B, V}, and {\it R}) data, \cite{de2006star} argue that a Toomre-Q criterion derived using the lower velocity dispersion ($\sim$\,4\,\kms) of the kinematically cool component is better at describing ongoing local star formation in NGC 6822 than the one with a velocity dispersion value of $\sim$\,6\,\kms\ derived from the conventional profile analysis.

For the profile analysis, \cite{de2006star} fit two-component Gaussian models to all the line profiles in batches, and choose a preferred model for each line profile which has a significantly lower reduced $\chi^{2}$ value. However, the model selection based on the $\chi^{2}$ minimization method often suffers from being trapped by local minima in the course of finding the global minima, particularly for low signal-to-noise ratio (S/N) line profiles (\citealt{de2006star}, \citealt{young2003star}). In addition, the fitting tends to be sensitive to initial estimates of model parameters.

To improve the parameter estimation and the model selection in the profile analysis, we make a practical application of a newly developed profile decomposition algorithm, the so-called {\sc baygaud}\footnote{\url{https://sites.google.com/view/baygaud}} (\citealt{oh2019robust}; \citealt{oh2022kinematic}). 
{\sc baygaud} which is based on a Bayesian analysis technique allows us to decompose a line profile into an optimal number of Gaussian components, circumventing the limited capability of the conventional profile analysis, particularly for non-Gaussian line profiles.

In this paper, we decompose all the line profiles of the ATCA H{\sc i} data cube of NGC 6822 (\citealt{de2000evidence}) with an optimal number of Gaussian components using {\sc baygaud}, and classify them as kinematically cold and warm H{\sc i} gas components with respect to their velocity dispersion values. In addition, we also classify them as bulk and non-bulk H{\sc i} gas motions which are either consistent or inconsistent with the global kinematics of the galaxy. We then correlate the $\Sigma_{\rm GAS}$ of the decomposed H{\sc i} gas components with their corresponding $\Sigma_{\rm SFR}$ of the galaxy on a pixel-by-pixel basis. For the $\Sigma_{\rm SFR}$, we use GALEX far-ultraviolet (\citealt{de2007galex}) and WISE\footnote{Wide-field Infrared Survey Explorer} MIR 22 $\mu$m data (\citealt{wright2010wide}) to trace $\Sigma_{\rm SFR}$. From the data and methods briefly described above, we examine the effect of gas kinematics on star formation in NGC 6822.

In Section~\ref{sec:sect2}, we present the H{\sc i}, FUV, and MIR data of NGC 6822, and the estimation of SFRs of NGC 6822 using FUV and MIR data. Section~\ref{sec:sect3} presents the application of {\sc baygaud} to the ATCA H{\sc i} data cube. The kinematic classification of the decomposed components is described in Section~\ref{sec:sect4}. In Section~\ref{sec:sect5}, we discuss the distribution of the classified H{\sc i} gas. Section~\ref{sec:sect6} describes the K-S star formation law at low gas densities. We summarize the main results of this paper in Section~\ref{sec:sect7}.

\section{Data}
\label{sec:sect2}
\subsection{The ATCA H{\sc i} 21cm data cube}
\label{sec:sect2.1}
The proximity of NGC 6822 as well as its extended H{\sc i} gas disk allow us to investigate the physical properties of the atomic ISM and their detailed interplay with star formation activities at GMC scales (\citealt{cannon2006nature}; \citealt{de2006stellar}; \citealt{de2006star}; \citealt{namumba2017h}). For the kinematic analysis of atomic ISM in NGC 6822, we use the H{\sc i} data cube obtained from observations with the Australia Telescope Com-pact Array (ATCA) which is presented in \citet{de2000evidence}. The large H{\sc i} gas disk of the galaxy with a major axis of $\sim$\,1\,$^{\circ}$ was observed through a mosaic of eight pointings with ATCA. The raw H{\sc i} visibility data are reduced, calibrated and imaged with the {\sc miriad}\footnote{\resizebox{0.95\hsize}{!}{Multichannel Image Reconstruction, Image Analysis and Display}} using a super-uniform weighting at various spatial (42.4\arcsec $\times$ 12.0\arcsec, 86.4\arcsec $\times$ 24.0\arcsec, 174.7\arcsec $\times$ 48.0\arcsec, and 349.4\arcsec $\times$ 96.0\arcsec) and spectral (0.8\,\kms \space and 1.6\,\kms) resolutions. We refer to \citet{de2000evidence} for the complete description of the ATCA observations and data reduction (see also \citealt{weldrake2003high}, \citealt{de2006stellar}, and \citealt{de2006star}). The physical and H{\sc i} properties of NGC 6822 are listed in Table~\ref{tab:galaxyproperty}.

\begin{figure*}
    \centering
    \includegraphics[width=0.8\textwidth]{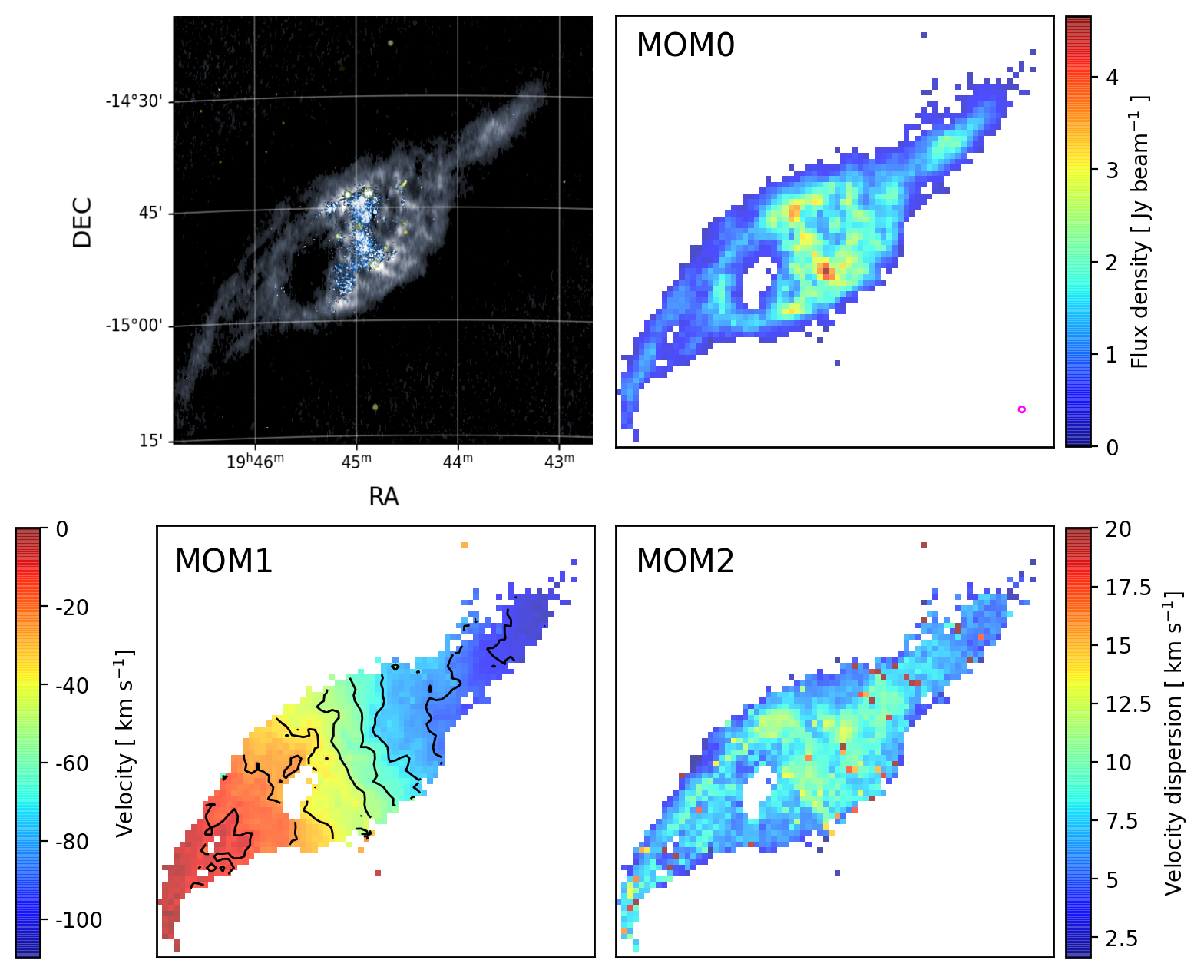}
    \caption{The top left panel shows a multi-wavelength color-composite image of NGC 6822 using the integrated flux density map from ATCA H{\sc i} 21cm (gray), GALEX FUV (blue), and WISE 22\,$\mu$m (yellow) data. The top right, bottom left, and bottom right panels show moment maps derived using beam sizes of 48.0\arcsec $\times$ 48.0\arcsec. This beam is shown in the bottom right corner (magenta) of the top right panel.}
    \label{fig:moment}
\end{figure*}

\setlength{\tabcolsep}{10pt}
\renewcommand{\arraystretch}{1}
\setlist[itemize]{leftmargin=5.5mm}

\begin{table}
    \centering
    \begin{tabular}{p{3.75cm}|p{2.1cm}|p{0.4cm}}
        \hline
        \hline
        Parameter       & Values    & Ref.\\
        \hline
        RA (J2000)      & 19h\,44m\,57s.74 & (a)   \\
        Dec (J2000)     & -14d\,48m\,12s.4 & (a)   \\
        Distance        & 490 $\pm \rm$ 40 kpc & (b) \\
        Metallicity     & 0.3 $Z_{\odot}$ &(c)   \\
        Position Angle (PA)  & 120\,$^{\circ}$ &(d)   \\
        Inclination ({\it i}\,)    & 60\,$^{\circ}$ &(d)    \\
        Systematic velocity (V$_{sys}$)       & $-55$ \,\kms  &(e)    \\
        Total H{\sc i} mass  & 1.3 $\times \rm$ $10^{8}$ M$_{\odot}$ & (e) \\
        Dynamical mass  & 4.3 $\times \rm$ $10^{9}$ M$_{\odot}$ & (e) \\
        H{\sc i} diameter    & 12 kpc & (e)   \\
        Optical diameter (D$_{25}$)    & 2.7 kpc & (e) \\
        \hline
        \hline
    \end{tabular}
    \caption{Physical properties of NGC 6822. The estimates are from (a) NED$^{a}$ (b) \citet{mateo1998dwarf}, (c) \citet{skillman1989abundances}, (d) \citet{weldrake2003high}, and (e) \citet{de2000evidence}.}
    \begin{itemize}
        \item [a] The NASA/IPAC Extragalactic Database (NED) is funded by the National Aeronautics and Space Administration and operated by the California Institute of Technology.
    \end{itemize}
    \label{tab:galaxyproperty}
\end{table}

In this work, we use the H{\sc i} data cube which is imaged with a beam size of 42.4\arcsec $\times$ 12.0\arcsec\ and a spectral resolution of 1.6\,\kms\ for the profile analysis of NGC 6822. We smooth the cube spatially using a 2D Gaussian kernel to make its final resolution be 48.0\arcsec $\times$ 48.0\arcsec\,per each pixel. This is to make individual line-of-sight (LOS) velocity profiles independent of each other. The resulting pixel scale of the cube is $\sim$\,114\,pc which matches with a typical size of GMC complex in galaxies (\citealt{sanders1985giant}). For the convolution and reformatting of the cube, we use the {\sc imsmooth} task in CASA\footnote{Common Astronomy Software Applications, Version 5.6.2} and the {\sc regrid} task in GIPSY{\footnote{Groningen Image Processing System}}, respectively.

The velocity range of the ATCA H{\sc i} data cube includes part of the velocity ranges ($-14$\,\kms\,$\leq$ v $\leq$ $-3$\,\kms) over which the Galactic H{\sc i} emission is significant, particularly in the southeastern (SE) region of NGC 6822 (\citealt{de2000evidence}, \citealt{de2006stellar}, \citealt{de2006star}; see also \citealt{namumba2017h}). Despite the fact that observations with radio interferometers tend to be less sensitive to the Galactic H{\sc i} emissions at large scales due to the lack of information about low spatial frequencies, parts of them are found to be still present in the ATCA data cube from visual inspection. We therefore replace the fluxes of the corresponding channels (i.e., $-14$\,\kms\,$\leq$ v $\leq$ $-3$\,\kms) with the ones derived from a linear interpolation using the adjacent channels ($-14$\,\kms\ and $-3$\,\kms) on the two velocity sides of the Galactic H{\sc i} emission.

We extract moment maps ({\sc moment0}: integrated flux density, {\sc moment1}: intensity-weighted mean velocity, and {\sc moment2}: intensity-weighted velocity dispersion) from the ATCA H{\sc i} data cube which is corrected for the Galactic H{\sc i} emission, using the {\sc moments} task in GIPSY. To minimise effect of any spike-like profiles on the moment analysis, we only consider the velocity channels whose adjacent fluxes are greater than 2 root-mean-square (rms) when deriving the moment values for each line profile. The extracted moment maps are presented in Fig.~\ref{fig:moment}, and the beam size is located in the bottom-right corner of top right panel. The lowest and highest inclination-corrected column densities derived using the H{\sc i} {\sc moment0} are $3.6 \times 10^{19}$ cm$^{-2}$ and $3.5 \times 10^{21}$ cm$^{-2}$, respectively, which are consistent with the ones in \citet{de2006stellar} (see also \citealt{namumba2017h}).

One notable feature in the H{\sc i} gas distribution of NGC 6822 ({\sc moment0}) is a companion H{\sc i} cloud which is located in the northwestern (NW) region of the galaxy at a galactocentric distance of $\sim$\, 4 kpc (\citealt{de2003young}; \citealt{de2006stellar}). \citet{de2003young} argue that the concentrated OB and blue stars in the region can be the results of star formation triggered by the interaction of NGC 6822 with the cloud.

Several H{\sc i} holes and shells are present in NGC 6822, and of which the supergiant H{\sc i} shell occupying a large area of the disk with a diameter of $\sim$ 1\,kpc in the SE region is the most outstanding H{\sc i} feature in the galaxy. Several studies discuss about the origin of supergiant shells in galaxies in the context of stellar wind, subsequent supernova explosions or an impact of infalling small high-velocity clouds (HVCs) (\citealt{de2000evidence}; \citealt{de2006stellar}; \citealt{kulkarni1988neutral}; \citealt{van1996bubbles}; \citealt{brinks1998violent}; \citealt{cannon2012origin}). Similarly, such radiative or mechanical impact might have created the supergiant H{\sc i} shell in NGC 6822. However, these are not obvious for the origin of the shell since no clear optical or kinematical signature of any leftover star clusters and HVC remnants is found near the hole as discussed in \citet{de2000evidence}.

\subsection{The GALEX {\it FUV} and WISE {\it MIR} data}
\label{sec:sect2.2}
We use the GALEX far-ultraviolet ({\it FUV}; 1350\---1750\,\AA; $\lambda_{e\!f\negmedspace f}\sim$\,1516\,\AA) and WISE mid-infrared ({\it MIR}; 22\,$\mu$m) Atlas images of NGC 6822 in order to estimate the star formation rates (SFRs) of the galaxy. The {\it FUV} and {\it MIR} bands have been widely used for tracing star formations in galaxies (e.g., \citealt{catalan2015star}; \citealt{bigiel2008star}; \citealt{leroy2008star}; \citealt{roychowdhury2015spatially}; \citealt{de2019revisiting}; \citealt{kennicutt2021revisiting}). 

Firstly, we use the archival{\footnote{Milkulski Archive for Space Telescope, MAST}}$^{,}$ {\footnote{\resizebox{0.95\hsize}{!}{\url{https://mast.stsci.edu/portal/Mashup/Clients/Mast/Portal.html}}}} GALEX {\it FUV} intensity map of NGC 6822 that is obtained as part of the GALEX Nearby Galaxies Survey (NGS; \citealt{de2007galex}). The spatial resolution of the {\it FUV} map is 4.2\arcsec $\times$ 4.2\arcsec, sampled at a pixel scale of 1.5\arcsec $\times$ 1.5\arcsec. For the full description of data and its unit conversion from [count\,sec$^{-1}$] to [erg\,sec$^{-1}$], we refer to the webpage of the GALEX data archive\footnote{\resizebox{0.95\hsize}{!}{\url{https://asd.gsfc.nasa.gov/archive/galex/Documents/ERO_data_description_2.htm}}}.

We construct a sky background map of the GALEX {\it FUV} image using a python package, `Photutils'\footnote{\url{https://photutils.readthedocs.io/en/stable/}} which applies a 3-$\sigma$ clipping method to the image and derives a median value of the background levels. We then subtract it from the GALEX {\it FUV} intensity map. In addition, we also remove foreground stars in the {\it FUV} image using the GALEX near-ultraviolet ({\it NUV}; 1750\---2800 \AA; $\lambda_{e\!f\negmedspace f} \sim$\,2267 \AA) image of NGC 6822 which is obtained from the GALEX NGS. For this, as described in \cite{bigiel2008star} and \cite{leroy2008star}, we replace the pixels in the {\it FUV} image whose {\it NUV}-to-{\it FUV} flux ratio $>$ 10 and {\it NUV} S/N $>$ 5 with the mean flux value of source-free regions.
 
Secondly, we use the WISE-W4 band (22 $\mu$m) Atlas image of NGC 6822 which is available from the WISE data archive (\citealt{wright2010wide}; \citealt{wise4band}; see also IRSA\footnote{\url{https://irsa.ipac.caltech.edu/}}). The WISE-W4 band is efficient for tracing dust attenuated star formation activity in the H{\sc i} gas disk of the galaxy (\citealt{catalan2015star}; \citealt{cluver2017calibrating}). A set of 115 images are combined to produce the WISE-W4 band image of NGC 6822 which covers the H{\sc i} gas disk of NGC 6822. The resolution of the WISE-W4 band Atlas image is 17\arcsec $\times$ 17\arcsec, and the pixel scale is 1.372\arcsec $\times$ 1.372\arcsec. 

Similarly, as for the reduction of the GALEX {\it FUV} image, we construct a sky background map of the WISE-W4 band image using the `Photutils' package, and subtract it from the WISE-W4 band image. For the removal of foreground stars in the WISE-W4 image, we use the {\it Spitzer} source catalog (Table 2 of \citealt{khan2015spitzer}) which provides the coordinates and 24 $\mu$m magnitudes of individual stellar components in NGC 6822. We identify foreground stars which are not listed in the catalogue but are bright enough in the WISE-W4 image. We replace the pixel values of the foreground stars with the mean flux value of source-free regions. The unit of the image, [DN] is converted into the luminosity in units of [erg\,sec$^{-1}$] using the method described in \citet{wright2010wide} (see also \citealt{jarrett2011spitzer}; \citealt{jarrett2013extending}, and IRSA Explanatory Supplement\footnote{\resizebox{0.9\hsize}{!}{\url{https://wise2.ipac.caltech.edu/docs/release/allsky/expsup/}}}).

Lastly, for the pixel-by-pixel comparison of the data sets, we match the spatial resolutions and sky coordinates of the three data sets, the ATCA H{\sc i} data cube, GALEX {\it FUV} and WISE {\it MIR} images by degrading the finer resolutions of the {\it FUV} and {\it MIR} images (FUV$-$4.2\arcsec\ and MIR$-$11.0\arcsec) to the one of the H{\sc i} data (48.0\arcsec). 
The point spread functions (PSFs) of the GALEX and WISE images are not Gaussian. As described in \citet{aniano2011common}, it can cause an artifact when converting the original image with a non-Gaussian PSF into the one assuming a Gaussian PSF. \citet{aniano2011common} generated various kernels to match the PSF of a given instrument with Gaussian PSF with numerous FWHM sizes. From these kernel sets, we use two kernels for the GALEX and WISE images, respectively, to mimic a common Gaussian PSF with FWHM of 20\arcsec. We refer to \citealt{aniano2011common} and a webpage\footnote{\url{https://www.astro.princeton.edu/~draine/Kernels.html}} for the full description of the kernel construction.

\begin{figure}
    \centering
    \includegraphics[width=0.9\columnwidth]{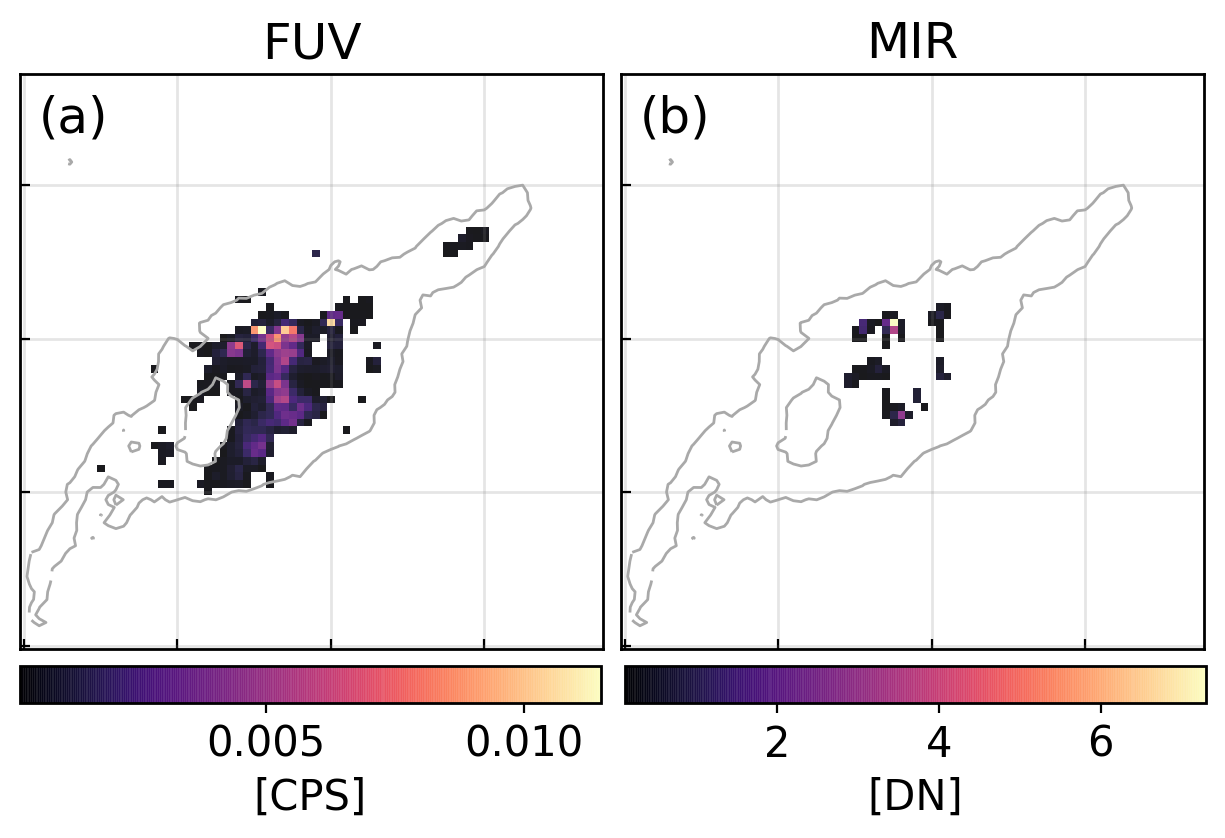}
    \caption{The GALEX {\it FUV} and WISE-W4 band images of NGC 6822 degraded into 48.0\arcsec $\times$ 48.0\arcsec. The contour presents H{\sc i} gas distribution where its column density is higher than $3.6 \times 10^{19}$ cm$^{-2}$.} 
    \label{fig:fuvmir}
\end{figure}

We then convolve the GALEX {\it FUV} and WISE {\it MIR} images with 20\arcsec\ FWHM using a 2D Gaussian kernel to make their final resolutions 48.0\arcsec $\times$ 48.0\arcsec, the same as H{\sc i} data. For this, we use the task {\sc imsmooth} in CASA. We then re-project the convolved images to match their coordinates and pixel scales with those of the H{\sc i} data cube using the {\sc reproj} task in GIPSY. In the course of the re-projection, we only keep the emissions with S/N larger than 4. The reduced GALEX {\it FUV} and WISE-W4 band images of NGC 6822 are presented in the panels (a) and (b) of Fig.~\ref{fig:fuvmir}.

\subsection{Estimation of star formation rates}
\label{sec:sect2.3}
We derive a hybrid SFR map of NGC 6822 using both the GALEX {\it FUV} and WISE {\it MIR} images following the method described in \citet{calzetti2013star} (see also \citealt{kennicutt2012star}). The {\it FUV} image is useful for tracing the unobscured recent star formation ($\leq$\,$\sim$\,100 Myr) in the galaxy as {\it FUV} photons are mainly produced by massive O and B type stars, and less contaminated by old stellar populations including the Galactic foreground stars (e.g., \citealt{bianchi2011galex}). However, interstellar dust can interact with the {\it FUV} photons through scattering and absorption processes, and re-radiate them at longer wavelengths mainly at {\it MIR} bands. To correct for the effect of such dust attenuation on the total SFR of a galaxy, the {\it MIR} image is usually combined into the unobscured FUV emission as described in several studies (e.g., \citealt{calzetti2007calibration}; \citealt{calzetti2005star}; \citealt{liu2011super}; \citealt{kennicutt2009dust}; \citealt{hao2011dust}; \citealt{catalan2015star}).

Firstly, for the region of NGC 6822 for which both the GALEX {\it FUV} and WISE {\it MIR} data are available, we use an empirical relation given in \citet{liu2011super} which derives a hybrid local SFR using the fluxes measured at {\it FUV} and 24\,$\mu$m {\it MIR} wavelengths as follows:
        
\begin{equation}
    \begin{split}
        \Sigma_{SFR(FUV,24)} = (9 \times 10^{-29} [L_{FUV, obs}\, ster^{-1} \\+ 0.0417 \times L_{24, obs}\, ster^{-1}])\,cos\,i
    \end{split}
    \label{eq:sfr_both}
\end{equation}

where {$L_{FUV, obs}$} and {$L_{24, obs}$} are the observed luminosities at the GALEX {\it FUV} and the {\it Spitzer}-MIPS 24\,$\mu$m in units of [erg sec$^{-1}$ Hz$^{-1}$], respectively, and {\it i} is the inclination of the galaxy. The derived $\Sigma_{SFR(FUV,24)}$ are in units of [M$_{\odot}$ yr$^{-1}$ kpc$^{-2}$]. This relation assumes the Kroupa initial mass function (IMF) (\citealt{kroupa2001on}) and no significant variation of SFR for the last 100 Myrs and fully sampled stellar IMF. 

We use the flux ratio between the WISE-W4 22\,$\mu$m and the {\it Spitzer}-MIPS 24\,$\mu$m to make use of 22\,$\mu$m luminosity as the substitute for the 24\,$\mu$m in Eq.~(\ref{eq:sfr_both}). In \citeauthor{jarrett2013extending} (\citeyear{jarrett2013extending}; see Table 5), the derived total flux ratio of WISE-W4 22\,$\mu$m to the {\it Spitzer} MIPS 24\,$\mu$m for NGC 6822 is 1.006. We multiply a factor of 0.994 to the WISE-W4 22\,$\mu$m luminosity when converting it to the $L(24)_{obs}$. We then plug the corresponding {\it FUV} and 24\,$\mu$m luminosity values of NGC 6822 into Eq.~(\ref{eq:sfr_both}).

Secondly, for the region of NGC 6822 for which only the GALEX {\it FUV} data is available, we use a model-based relation given in \citeauthor{calzetti2013star} (\citeyear{calzetti2013star}; see Eq. 1.2 and Table 1.1) to derive the unobscured SFR as follows:

\begin{equation}
    \begin{split}
        \Sigma_{SFR(FUV)} = (9 \times 10^{-29} [L_{FUV, corr}\, ster^{-1}])\,cos\,i
    \end{split}
    \label{eq:sfr_fuvonly}
\end{equation}

where $L_{FUV, corr}$ is the dust extinction corrected luminosity at the GALEX {\it FUV} in units of [erg sec$^{-1}$ Hz$^{-1}$], {\it i} is the inclination of the galaxy. The derived $\Sigma_{SFR(FUV)}$ is in units of [M$_{\odot}$ yr$^{-1}$ kpc$^{-2}$].

We correct for the extinction caused by the foreground dust in the Milky Way (MW) and NGC 6822 internally, using the standard MW extinction curve with $A_{FUV} = 8.24$ $\times$ $E(B-V)_{t}$ and $R_{V} = 3.1$ as described in \cite{cardelli1989relationship} where $E(B-V)_{t}$ is the sum of foreground stellar color excess, $E(B-V)_{f}$ and the internal color excess, $E(B-V)_{int}$. \citet{efremova2011recent} derived individual $E(B-V)_{t}$ values for 77 star forming regions which mainly locate the inner region of NGC 6822 using CTIO {\it U}, {\it B}, and {\it V} images. The derived $E(B-V)_{t}$ range from 0.22 (MW extinction only) to 0.66 mag with a mean value of 0.36, which is higher than $E(B-V)_{t} = 0.27$ presented in \citet{hunter2010galex}. We adopt the mean value of $E(B-V)_{t} = 0.37$ when deriving $A_{FUV}$ which is used for deriving the dust extinction corrected luminosity $L_{FUV, corr}$ in Eq.~(\ref{eq:sfr_fuvonly}). The resulting total SFR of NGC 6822 is 0.015 M$_{\odot}$\,yr$^{-1}$ which is well consistent with the previous ones 0.012 M$_{\odot}$\,yr$^{-1}$ and 0.014 M$_{\odot}$\,yr$^{-1}$ derived in \citet{hunter2010galex} and \citet{efremova2011recent}, respectively.

\section{Profile decomposition}
\label{sec:sect3}
\subsection{HI profile decomposition}
\label{sec:sect3.1}
We model individual line profiles of the spatially smoothed ATCA H{\sc i} data cube (48.0\arcsec $\times$ 48.0\arcsec) described in Section~\ref{sec:sect2} with multiple Gaussian components using a profile decomposition tool, {\sc baygaud} (\citealt{oh2019robust}). From the test described in \citet{oh2019robust}, {\sc baygaud} is found to be robust for parameter estimation and model selection against any local minima when decomposing a non-Gaussian velocity profile into an optimal number of Gaussians as long as it has sufficient signal-to-noise (S/N), for example, larger than 3 or so. We refer to \citet{oh2019robust} for the full description of the fitting algorithm and its performance test.

For each line profile, {\sc baygaud} performs profile decomposition with an optimal number of Gaussian components based on Bayesian analysis techniques. Most line profiles are non-Gaussian, and from a visual inspection of the line profiles of the cube, modelling the profiles with three Gaussians is likely to be enough to take their non-Gaussianity into account. Considering the non-Gaussianity of the profiles, we let {\sc baygaud} fit a line profile with up to three Gaussian components simultaneously. For instance, it makes three Bayesian fits to each line profile with $m$=1 to 3 using the following equation,
\begin{equation}
    \label{eq:baygaud}
    G(v) = \sum_{i=1}^{m} \frac{A_{i}}{\sigma_{i}\sqrt{2\pi}} \exp\bigg(\frac{-(v-\mu_{i})^{2}}{2\sigma_{i}^{2}}\bigg) + \sum_{j=0}^{n} b_{j}v^{j}\,,
\end{equation}
where G($v$) is an velocity profile, {\it m} is the maximum number of Gaussian components, and $A_{i}$, $\sigma_{i}$, and $\mu_{i}$ are the Gaussian parameters of integrated intensity, velocity dispersion, and central velocity of the {\it i}-th Gaussian component, respectively. $b_{j}$ are coefficients of the $n^{th}$ order polynomial function which is for modelling the baseline. In this work, we set $n = 1$ for the baseline fitting.

Among the three different models, {\sc baygaud} then finds the most appropriate Gaussian model for each line profile given their Bayesian evidence which is derived using {\sc multinest} (\citealt{feroz2008multimodal}), a Bayesian inference library. Practically, it uses Bayes factor which is the prior probability ratio of two competing models. In this work, we use the so-called 'substantial' model selection criteria in which one model is preferred to the other if its Bayes factor is greater than 3.2 (\citealt{jeffreys1998theory}; \citealt{raftery1995bayesian}). In this way, we adopt the best model whose Bayes factor against the second best model is at least larger than 3.2. We refer to \cite{oh2019robust} for more details.

In the rest of this paper, we designate the Bayesian profile decomposition with multiple Gaussian components as `OptGfit'. Together, we also fit a single Gaussian function to the line profiles of the cube, which provides 2D maps of integrated flux density, central velocity and velocity dispersion. We use the Bayesian analysis for this single Gaussian fitting. We call this single Gaussian fitting analysis as `SGfit'.

\subsection{The fit quality of Gaussian models}
\label{sec:sect3.2}
\begin{figure}
    \centering
    \includegraphics[width=1.0\columnwidth]{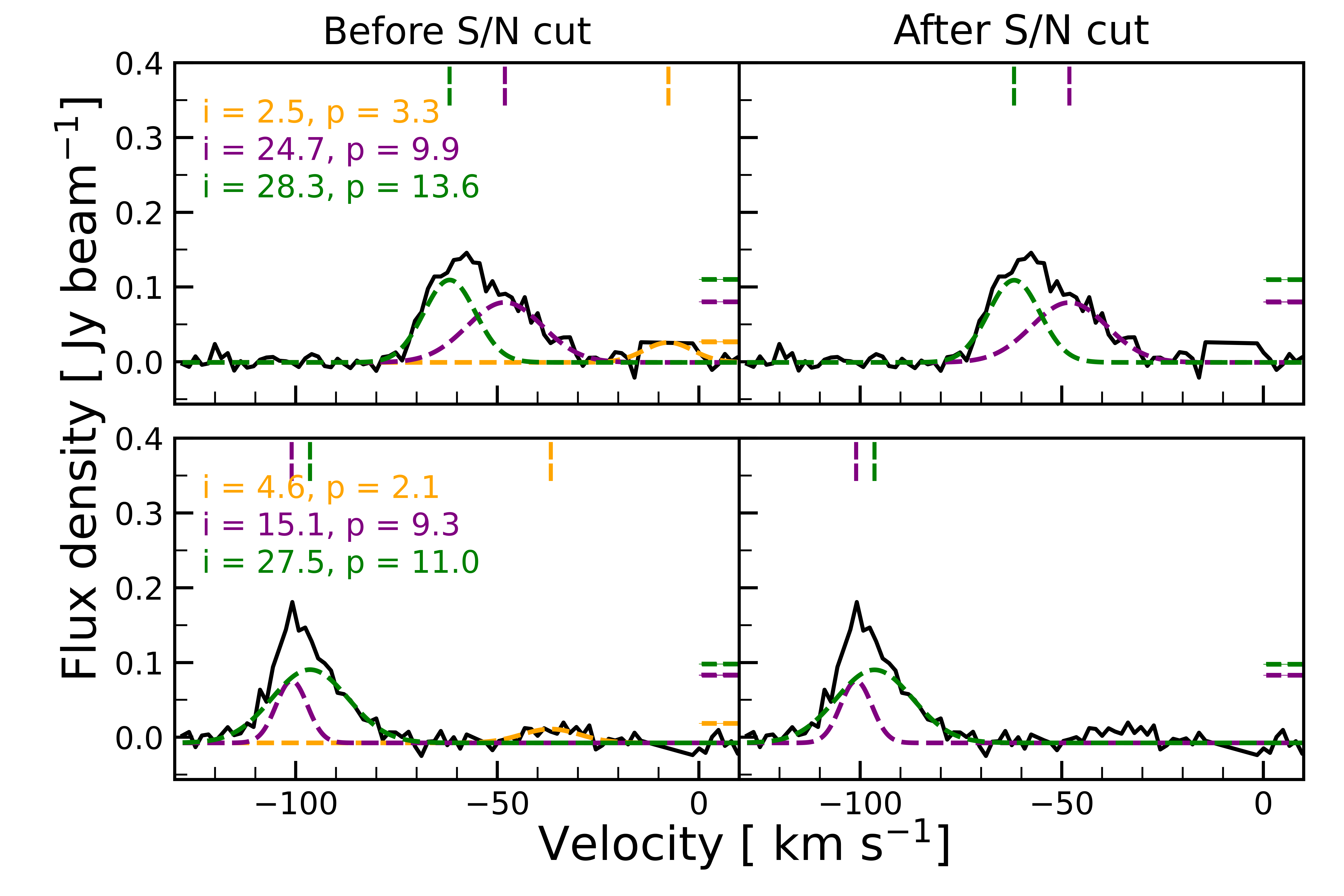}
    \caption{Two example profiles (black solid lines) showing the S/N cut adopted in this work. The dashed lines in the left panels show the OptGfit results, and their integrated (`i') and peak (`p') S/N values are denoted on the top-left corner of each panel. The central velocities and flux density values are indicated by the vertical and horizontal lines on the top and right sides of each panel, respectively. The resulting Gaussian components obtained after filtering out the ones with low S/N are shown in the right panels.}
    \label{fig:sncut}
\end{figure}

We reject the decomposed Gaussian components whose `integrated' S/N value is lower than 4. The integrated S/N cut is efficient for removing any spike-like profiles (\citealt{popping2012comparison}; \citealt{meyer2017tracing}; \citealt{for2019wallaby}). We derive integrated rms of a line profile as follows,
\begin{equation}
    rms_{int} = \sqrt{n_{chan}} \times rms_{bg}\times \delta V,
    \label{eq:integratedsn}
\end{equation}
\noindent where $n_{chan}$ is the number of channels available for a given Gaussian component, and $rms_{bg}$ is the estimated rms of the background level of the profile measured from {\sc baygaud}. $n_{chan}$ is computed by $6 \times v_{di\!sp} / \delta V$ where $v_{di\!sp}$ is the velocity dispersion obtained from SGfit, and $\delta V$ is the channel width of the cube. Over the velocity range corresponding to $n_{chan}$, $\sim$\,99.7\% of the integrated flux of a line profile is covered. The top panels of Fig.~\ref{fig:sncut} show an example profile in which one component (yellow) on the left panel is filtered out due to its integrated S/N value lower than 4.

In addition, we also exclude Gaussian components whose peak S/N value is lower than 3 regardless of its integrated S/N value. The bottom panels of Fig.~\ref{fig:sncut} describe that one of three Gaussian components (yellow) on the left panel is filtered in the final OptGfit result, since its peak S/N of 2.1. We substitute the SGfit results of a line profile into its OptGfit ones unless all the decomposed Gaussian components of the profile pass through the profile rejection criteria adopted in this work. There are some profiles which are decomposed with multiple Gaussian components but have a significantly lower background level than their ambient profiles. This could be due to background ripples which can be caused by low-order baseline fitting or strong absorption features in the profiles. Like the case of line profiles with low S/N, for the profiles whose background level is lower than $bg_{med} - 1\times rms_{bg}$ where $bg_{med}$ is the median value of background levels of all the line profiles, we replace their OptGfit results with the SGfit ones.

\begin{figure*}
    \centering
    \includegraphics[width=0.7\textwidth]{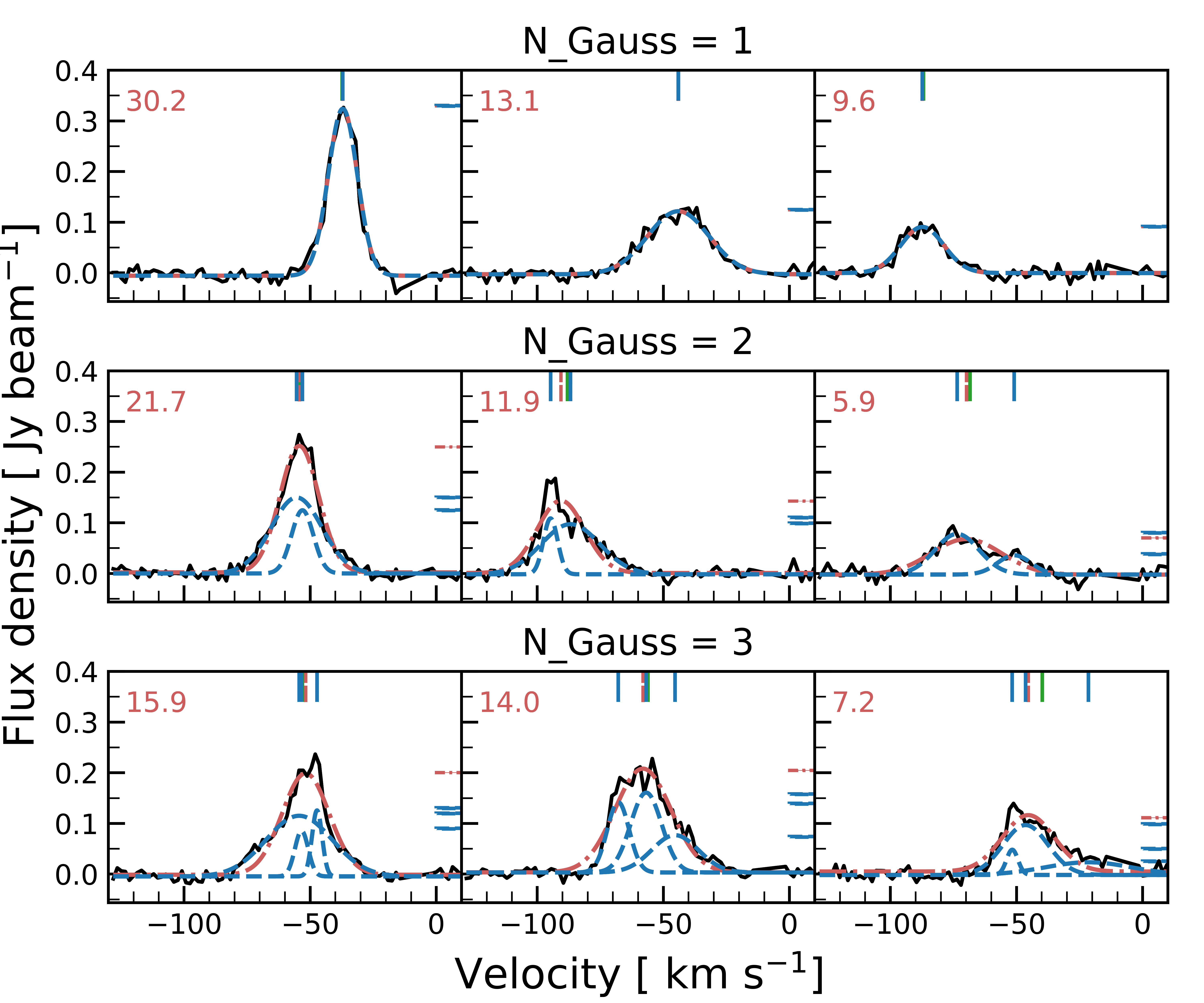}
    \caption{Arbitrarily chosen nine velocity profiles (black solid lines). From the top to the bottom, the optimal number of Gaussian in describing the individual line profile is 1, 2, and 3, respectively. The left, middle, and right columns present velocity profiles whose peak S/N estimated from SGfit is larger than 15, between 15 and 10, between 10 and 5, respectively. The red dash-dotted and blue dashed lines present the SGfit and OptGfit results, respectively. The central velocity measured by {\sc moments} (green solid), SGfit (red dash-dotted), and OptGfit (blue dashed) is indicated as vertical lines on the top of each panel. The flux densities derived from SGfit and OptGfit are also shown as horizontal lines on the right of each panel. The estimated peak S/N value of SGfit result is shown on the top-left corner of each panel.}
    \label{fig:example}
\end{figure*}

\begin{figure}
    \centering
    \includegraphics[width=0.7\columnwidth]{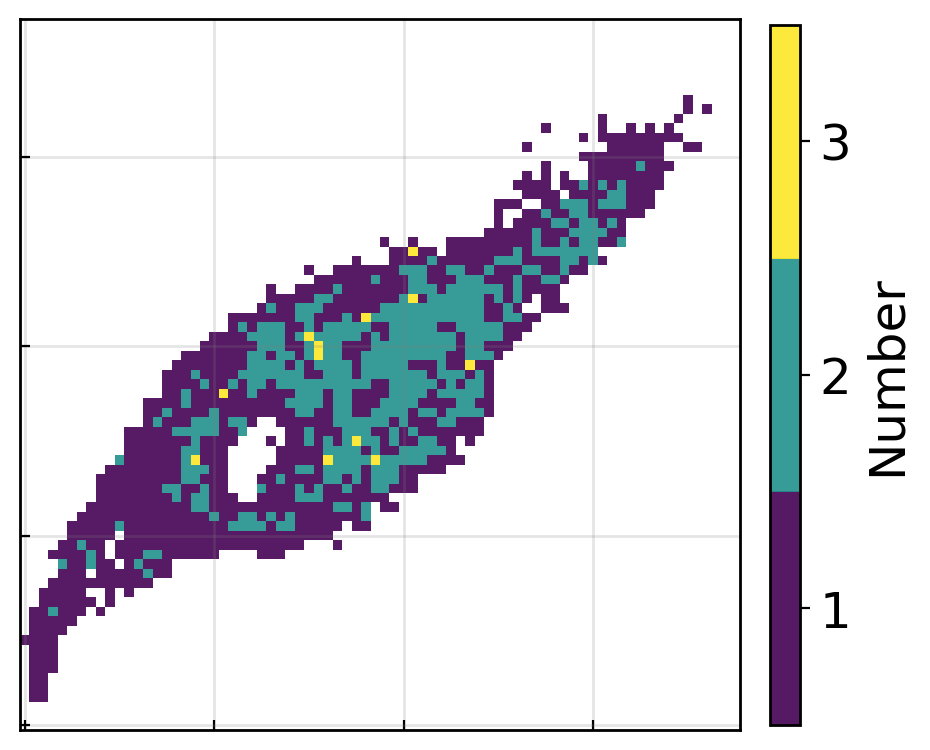}
    \caption{The number of optimally decomposed Gaussian map. Blue, green, and yellow points present the optimal number of Gaussian is one, two, and three, respectively.}
    \label{fig:ngauss}
\end{figure}

Example line profiles (black solid lines) of the ATCA H{\sc i} data cube and their profile decomposition (dashed lines) are shown in Fig.~\ref{fig:example}. The blue and brown dashed lines represent the OptGfit and SGfit results, respectively. Compared to the SGfit results, the OptGfit ones better model non-Gaussian line profiles with an optimal number of multiple Gaussian components, retrieving more accurate fluxes of the profiles. This is particularly pronounced for the profiles which can be modelled by double or triple Gaussian components. Fig.~\ref{fig:ngauss} shows a map of the number of Gaussian components which are decomposed from the ATCA H{\sc i} data cube of NGC 6822 in this work. The non-Gaussian shape of H{\sc i} line profiles of the galaxy is described by the number of Gaussian components with two or even three in some regions. This allows us to understand the complex structure and kinematics of the H{\sc i} gas component in the galaxy in a quantitative manner.

\begin{figure}
    \centering
    \includegraphics[width=0.87\columnwidth]{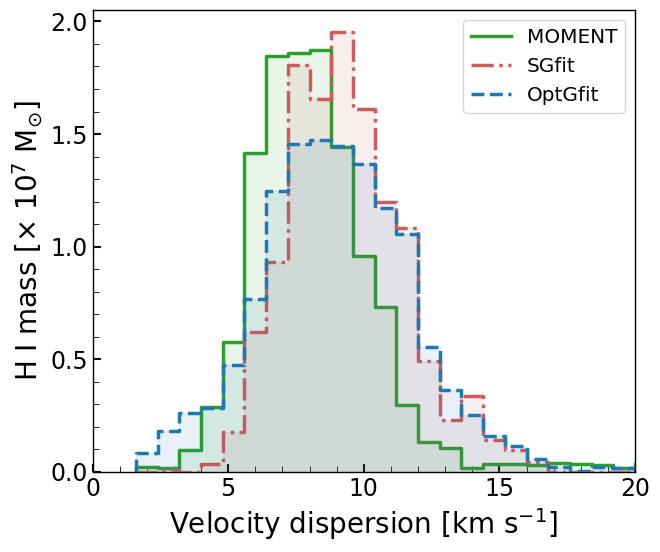}
    \caption{The mass distribution for the velocity dispersions. The bin width of the histogram is 0.8\,\kms. Green, red, and blue bars indicate the H{\sc i} mass distribution derived from {\sc moment0}, SGfit, and OptGfit ({\sc baygaud}) integrated flux density maps, respectively.}
    \label{fig:veldisp_mass}
\end{figure}

\subsection{Estimation of HI mass}
\label{sec:sect3.3}
We derive the H{\sc i} masses of NGC 6822 which correspond to the integrated flux densities derived using the SGfit and OptGfit results as well as the {\sc moment0}. As described in Section~\ref{sec:sect2.1}, when deriving the {\sc moment0} map from the cube, we only include the velocity channels whose adjacent channel fluxes are greater than two rms. This is for excluding any spike-like, one channel wide noise profile. It is then masked to remove the pixels whose integrated S/N is lower than four as in the cases of the SGfit and OptGfit analysis. The H{\sc i} integrated flux densities of NGC 6822, $S^{HI}_{int}$ which is in units of Jy\,\kms, and measured from the three methods are plugged into the following equation to derive the corresponding H{\sc i} masses (\citealt{roberts1962neutral}):
\begin{equation}
    \left(\frac{M_{HI}}{M_{\odot}}\right) = 0.236 \times 10^{5} \left(\frac{D}{kpc} \right)^{2} \left(\frac{S^{HI}_{int}}{Jy\,\kms}\right),
    \label{eq:himass}
\end{equation}
where $D$ is the distance to NGC 6822, 490\,kpc.

\begin{table}
    \centering
    \begin{tabular}{c|c}
        \hline
        \hline
        Method & H{\sc i} mass [$\times 1$E8\,M$_\odot$] \\
        \hline
        Moment & 1.20 \\
        SGfit & 1.24 \\
        OptGfit & 1.29 \\
        \hline
        \hline
        \end{tabular}
    \caption{Total H{\sc i} mass of NGC 6822 derived from {\sc moment0}, SGfit, and OptGfit.}
    \label{tab:total_mass}
\end{table}

The H{\sc i} masses derived from {\sc moment0}, SGfit and OptGfit are listed in Table~\ref{tab:total_mass}. The H{\sc i} mass derived using the $S^{HI}_{int}$ of the OptGfit is $\sim$7\% higher than the one derived from the moment analysis. This can be caused by the way of deriving the integrated flux densities of the profiles. As described in Section \ref{sec:intro} and \ref{sec:sect3.3}, we use 2$\sigma$\,\--clipping cut when constructing {\sc moment0} map. On the other hand, the integrated flux of a Gaussian profile model in OptGfit is calculated from -3\,$\sigma$ to +3\,$\sigma$ in the velocity axis, which covers $\sim$\,99\% of the total flux density.

Fig.~\ref{fig:veldisp_mass} shows the H{\sc i} mass distribution of NGC 6822 with respect to their corresponding velocity dispersions derived using the three different methods, {\sc moment0} (green dashed line), SGfit (red dash-dotted line), and OptGfit (blue solid line). This demonstrates that the OptGfit analysis retrieves more H{\sc i} mass in the low velocity dispersion bins ($v_{\rm disp}$\,$<$\,$\sim$\,4\,\kms) while less mass in the intermediate ones compared to both the moment and SGfit analyses. This is because of the nature of the OptGfit which models a line profile with multiple Gaussian components.

\section{Kinematic classification of H{\sc i} velocity profiles}
\label{sec:sect4}
\begin{figure}
    \centering
    \includegraphics[width=\columnwidth]{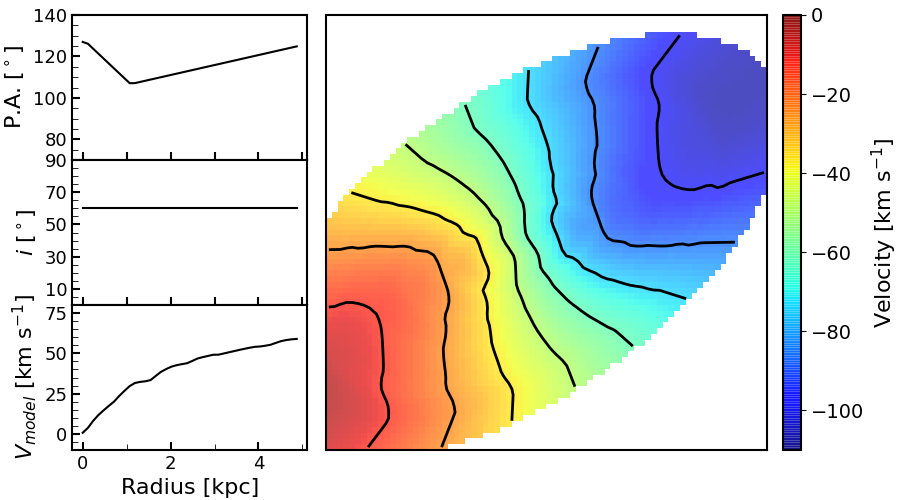}
    \caption{The model tilted-ring fit result derived in \citet{weldrake2003high}. The left panels show the radial model position angle (P.A), inclination ({\it i}), and model reference velocity (V$_{model}$). The right panel describes the model velocity field using the radial parameters on the left panel. The solid contours are in 10 \kms\,steps from $-$110 to 0 \kms.}
    \label{fig:modelvelfi}
\end{figure}

Of the decomposed Gaussian components of the H{\sc i} line profiles in Section~\ref{sec:sect3}, we extract H{\sc i} bulk motions which move at the velocities to the global kinematics of the H{\sc i} gas disk of NGC 6822 within a velocity limit. The global kinematics of the galaxy can be derived from the 2D tilted-ring analysis results presented in \citet{weldrake2003high}. A model reference velocity field for the global kinematics (V$_{model}$) derived using the {\sc velfi} task in GIPSY is shown in the bottom left panel of Fig.~\ref{fig:modelvelfi}. For the velocity limit, we use the velocity dispersions of individual line profiles which are derived from the SGfit analysis. Several hydrodynamical and gravitational effects in NGC 6822 including the gravitational potential and stellar feedback can give rise to the velocity dispersions of the H{\sc i} gas disk. In this work, we adopt the 2D map of the SGfit velocity dispersions as a conservative choice of velocity limit which covers the velocities of co-rotating gas motions associated with the global kinematics of the galaxy.

We classify the rest of the Gaussian components, deviating from the {\it bulk} motions as {\it non-bulk} motions. These non-bulk H{\sc i} gas motions can be attributed to hydrodynamical processes in NGC 6822. These include any gas motions which do not follow the global kinematics of the galaxy such as random, non-circular, and streaming motions etc. These can be associated with star formation activities in the galaxy. Together, its gravitational interactions with the Milky Way and/or a companion galaxy could partially contribute to the non-bulk gas motions. As discussed in \citet{zhang2021panoramic}, NGC 6822 might have encountered the Milky Way within its virial radius about 3 to 4 Gyrs ago.

\begin{figure}
    \centering
    \includegraphics[width=0.89\columnwidth]{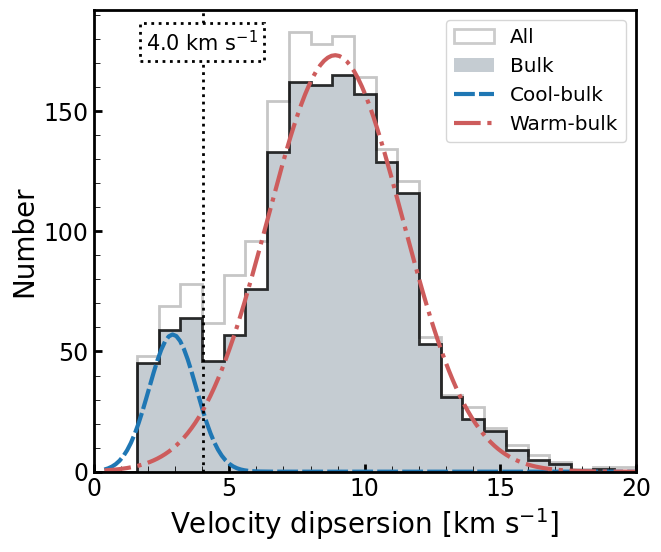}
    \caption{Velocity dispersion histogram extracted from the optimally decomposed Gaussian components (open histogram). The gray-shaded histogram is that for the bulk components. Two Gaussians describe the bimodality of the histogram and they overlap at 4.0\,\kms (black dotted line). Blue dashed line presents cool-bulk components ($v_{disp}$ $<$ 4.0\,\kms) and red dash-dotted line shows the warm-bulk distribution ($v_{disp}$ $>$ 4.0\,\kms).}
    \label{fig:veldisp_hist}
\end{figure}

Subsequently, we group the H{\sc i} bulk motions into cool-bulk and warm-bulk components with respect to their velocity dispersions, $v_{\rm disp}$. To set a boundary value of velocity dispersion between the two sub-components, we examine the histogram of velocity dispersions of the H{\sc i} bulk motions as shown in Fig.~\ref{fig:veldisp_hist}. A bimodal distribution of velocity dispersion is evident for the bulk components. We fit a double Gaussian model to the distribution, and determine a velocity dispersion value of $\sim$\,4 \kms\ at which the two Gaussian components are overlapped as indicated by the vertical dotted-line in Fig.~\ref{fig:veldisp_hist}. This value is similar to the one which is adopted for separating the H{\sc i} line profiles of NGC 6822 into `kinematically' cool and warm H{\sc i} components in \citet{de2006star}. 

Likewise, we also group the non-bulk H{\sc i} gas components into narrow and broad ones using their velocity dispersion. Since the non-bulk sample is not large enough to statistically decompose its velocity dispersion distribution, we apply the same velocity dispersion criteria (4 \kms) used for the bulk motions. By doing so, we classify the non-bulk H{\sc i} gas components as cool-non-bulk ($v_{\rm disp}$ $<$ 4 \kms) and warm-non-bulk ones ($v_{\rm disp}$ $>$ 4 \kms).

The mean velocity dispersion values of the cool-bulk, warm-bulk, cool-non-bulk, and warm-non-bulk components are $\sim$ 2.9 $\pm$ 0.7 \kms, $\sim$ 9.1 $\pm$ 2.5 \kms\, $\sim$ 3.2 $\pm$ 0.6\,\kms, and $\sim$ 8.1 $\pm$ 3.2 \kms, respectively. The cool-bulk component which is consistent with the global kinematics of the galaxy but kinematically colder than the warm-bulk component might be directly associated with gas reservoir for star formation as atomic hydrogen should pass through this cold phase in the course of forming molecular hydrogen (\citealt{ianjamasimanana2012shapes}). While the warm-bulk component might be caused by the radiative and mechanical stellar feedback in NGC 6822. The energy deposited from star formation in a galaxy can disturb its ambient ISM, which induces a larger velocity dispersion of the ISM. Or it can make the ISM deviated from the global kinematics of the gas disk if it is significant. This could be the case of the cool-non-bulk or warm-non-bulk H{\sc i} component. We derive the masses of the cool-bulk, warm-bulk, cool-non-bulk, and warm-non-bulk H{\sc i} components as listed in Table~\ref{tab:group_mass}. Their corresponding masses occupy $\sim\,3.8$\,$\%$, $\sim\,88.4$\,$\%$, $\sim\,0.8$\,$\%$, and $\sim\,7.0$\,$\%$ of the total H{\sc i} mass derived from the OptGfit analysis, respectively.

\begin{figure*}
    \vspace*{2cm}
    \centering
    \includegraphics[scale=0.65]{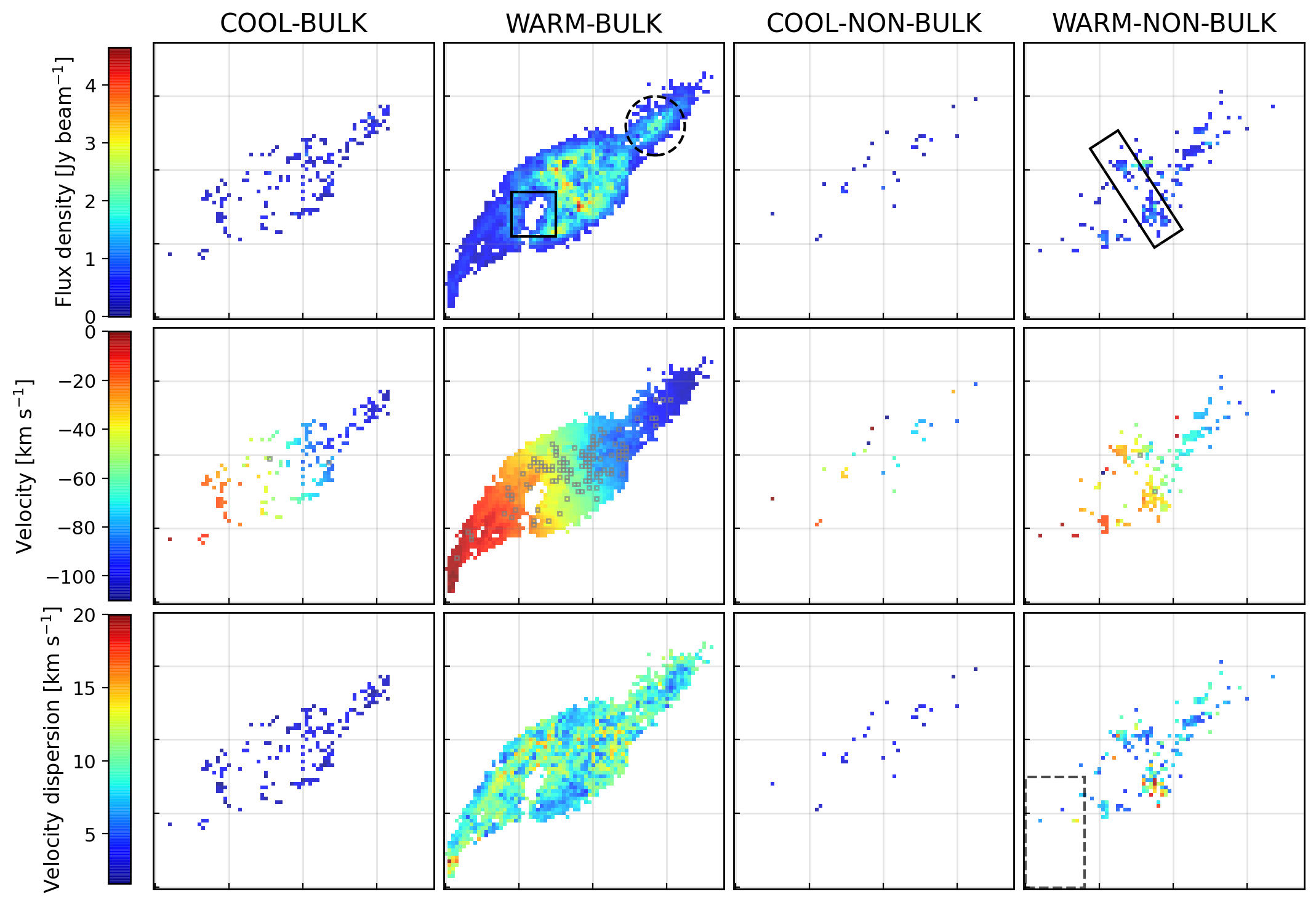}
    \caption{Atlas of kinematic classification of H{\sc i} gas. The figure shows the flux density, centroid velocity, and velocity dispersion maps of cool-bulk (V$_{disp} \leq $ 4.0\,\kms \space with bulk motion), warm-bulk (V$_{disp} > $ 4.0\,\kms \space with bulk motion), cool-non-bulk (V$_{disp} \leq $ 4.0\,\kms \space with non-bulk motion), and warm-non-bulk (V$_{disp} > $ 4.0\,\kms \space with non-bulk motion) gas components. The dashed circle on the warm-bulk flux density map represent the NW cloud. The solid square on the warm-bulk flux density map indicates the H{\sc i} shell, while the solid rectangular shows the possible candidate of HVCs. The points with gray empty boxes on each velocity map present the pixels having more than one components in the same classification. Black dashed rectangular on the rightmost-bottom panel show the regions we disregard in the following analysis.}
    \label{fig:atlas}
\end{figure*}

\begin{table}
    \centering
    \begin{tabular}{c|c}
        \hline
        \hline
        Kinematic classification & H{\sc i} mass [$\times$\,1E8 M$_\odot$]   \\
        \hline
        cool-bulk & 0.05   \\
        warm-bulk & 1.14   \\
        cool-non-bulk & 0.01   \\
        warm-non-bulk & 0.09  \\
        \hline
        \hline
        \end{tabular}
    \caption{The derived H{\sc i} mass of the cool-bulk, warm-bulk, cool-non-bulk, and warm-non-bulk gas components.}
    \label{tab:group_mass}
\end{table}

\section{The distribution of bulk and non-bulk H{\sc i} gas components}
\label{sec:sect5}
The extracted cool-bulk, warm-bulk, cool-non-bulk, and warm-non-bulk H{\sc i} components of NGC 6822 are mapped in Fig.~\ref{fig:atlas} in their flux density, central velocity and velocity dispersion. As shown in the figure, the H{\sc i} warm-bulk component is distributed throughout the entire disk of the galaxy. Some of H{\sc i} cool-bulk components appear to be in the vicinity of a companion H{\sc i} cloud and the supergiant shell in the NW and SE parts, respectively. The fraction of cool-non-bulk components is relatively small in the disk, showing a rather sporadic distribution. The shape of the H{\sc i} line profiles in the regions where the H{\sc i} cool-bulk or non-bulk components locates is mostly non-Gaussians In Table~\ref{tab:non_gaussian_fraction}, we list the mean number of Gaussian components of pixels where each kinematic component is present. For the pixels where two or three gas components are classified into the same kinematic type, we use the sum of their flux densities, and the mean values of their central velocities and velocity dispersions. These pixels are indicated by gray boxes in the velocity maps of Fig.~\ref{fig:atlas}.

\begin{table*}[]
    \centering
    \begin{tabular}{c|c}
    \hline
    \hline
    Kinematic classification & Mean number of Gaussian components \\
    \hline
    cool-bulk & 2.02 \\
    warm-bulk & 1.36 \\
    cool-non-bulk & 2.07 \\
    warm-non-bulk & 1.88 \\
    \hline
    \hline
    \end{tabular}
    \caption{Mean number of Gaussian components of pixels where each kinematic component is present. See Section~\ref{sec:sect5} for details.}
    \label{tab:non_gaussian_fraction}
\end{table*}

Hydrodynamical processes in NGC 6822, its gravitational interactions with other galaxies or both could stir up the H{\sc i} gas components in ways of 1) broadening the velocity width of line profiles (e.g., the warm-bulk motions and some of the non-bulk motions whose velocity dispersion is high), 2) deviating the gas motions from the global kinematics (e.g., the cool-non-bulk or warm-non-bulk motions) or 3) both (e.g., the warm-non-bulk motions whose velocity dispersion is high). They might enhance gas density, and thus trigger subsequent star formation as traced by the FUV emission in Fig.~\ref{fig:fuvmir}. On the other hand, undisturbed kinematically cold H{\sc i} gas components, namely cool-bulk components which co-rotate with the global kinematics but have low velocity dispersion managed to remain alongside the warm-bulk and non-bulk motions.

As mentioned in Section~\ref{sec:sect2.1}, the NW H{\sc i} cloud is prominent in the integrated flux density map of the warm-bulk component as denoted by a black dashed circle in Fig.~\ref{fig:atlas}. The cool-bulk H{\sc i} component might be directly associated with this recent star formation. On the other hand, both gravitational interaction and stellar feedback might have disturbed the non-bulk H{\sc i} gas component in the vicinity of the cloud as denoted by a circle in the top rightmost panel of Fig.~\ref{fig:atlas}.  

A notable H{\sc i} characteristic, the supergiant H{\sc i} shell in the SE region is indicated by a black solid square in the panel for the warm-bulk integrated flux density of Fig.~\ref{fig:atlas}. This feature is also seen in the integrated flux density maps of both the cool-bulk and warm-non-bulk motions although it is not evident as in that of the warm-bulk component. Parts of both the cool-bulk and warm-non-bulk components appear to be located along the edge of the shell. They could be affected by the shell in the course of its formation and evolution.

Another noticeable feature in the distribution of warm-non-bulk H{\sc i} component is indicated by a solid rectangular in its flux density map of Fig.~\ref{fig:atlas}. They are mainly located along the minor axis of the H{\sc i} disk which is determined from the 2D tilted-ring analysis in \citet{weldrake2003high}. These warm-non-bulk H{\sc i} gas motions can be HVC candidates as identified by the analysis of position-velocity diagram in \citet{de2006star}. The mean velocity offset of these warm-non-bulk motions from the global kinematics of the galaxy which is derived from the tilted-ring analysis is $\sim$\,14.6 \kms, while the mean value of their velocity dispersions is $\sim$\,8.9 \kms.

\section{The resolved Kennicutt-Schmidt law in NGC 6822}
\label{sec:sect6}

\begin{figure*}
    \centering
    \includegraphics[width=0.8\textwidth]{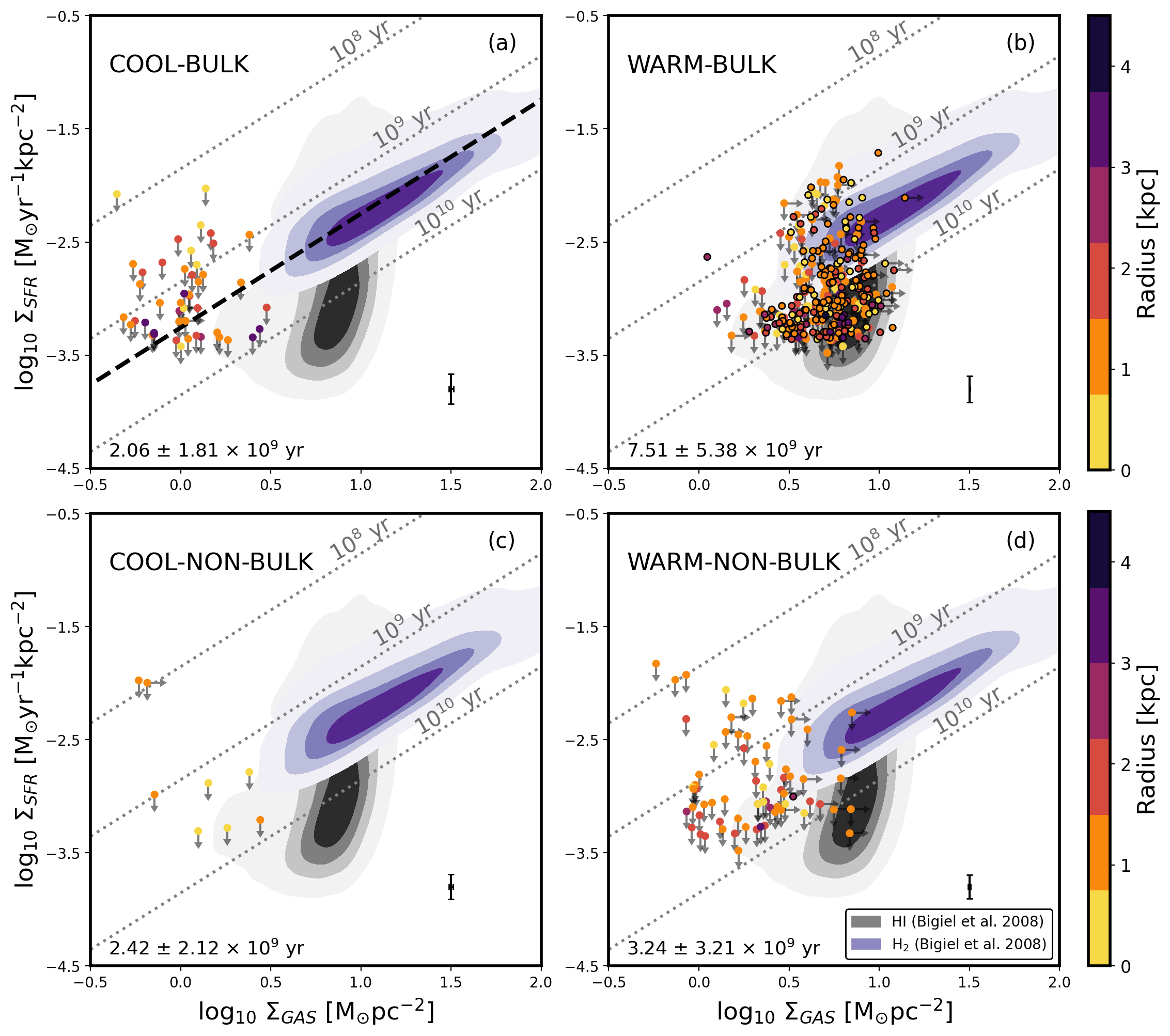}
    \caption{{\bf panels (a) to (d)}: The K-S law for gas components of each group varying with its galactocentric radius from 0 to 4.5\,kpc. The panels represent the resolved K-S law for the cool-bulk, warm-bulk, cool-non-bulk and warm-non-bulk components onto the K-S law of H{\sc i} (gray) and H$_{2}$ (blue) gas investigated in \citet{bigiel2008star}. The points with upper limits of $\Sigma_{SFR}$ are the gas components where the velocity profile is decomposed into more than one Gaussian component. The points with black edges present gas components whose velocity profile is decomposed into one Gaussian (i.e., SGfit) or have more two or three Gaussians, but all of them are classified into the same group. The points with lower limits of $\Sigma_{GAS}$ present regions where H$_{2}$ gas clouds are detected by previous studies. The dotted lines indicate the gas depletion time ($\tau_{dep}$), by which the gas consumption form stars in 10$^{8}$, 10$^{9}$, and 10$^{10}$ yr. The {\it x}-axis and {\it y}-axis error bars located in the in the lower right of each panel indicates the background uncertainties estimated from H{\sc i} and combination of FUV and MIR images, respectively. The x-axis error is 0.013, 0.003, 0.012, and 0.007 dex for the panels (a), (b), (c), and (d), respectively. The mean $\tau_{dep}$ of classified components is denoted on the bottom-left corner of each panel.}
    \label{fig:ks}
\end{figure*}

In this section, we investigate the relationship between the gas surface density ($\Sigma_{\rm GAS}$) and the SFR surface density ($\Sigma_{\rm SFR}$) in NGC 6822. As presented in \citeauthor{bigiel2008star} (\citeyear{bigiel2008star}; \citeyear{bigiel2010extremely}), Fig.~\ref{fig:ks} shows the $\Sigma_{\rm SFR}$ of THINGS spiral and dwarf galaxies against their corresponding $\Sigma_{\rm HI}$ or $\Sigma_{\rm H2}$ at a pixel scale of 750 pc. The abscissa and ordinate of the figure indicate the face-on (inclination corrected) logarithmic $\Sigma_{\rm GAS}$ of either H{\sc i} or $\rm H_{2}$ in units of [M$_{\odot}$\,pc$^{-2}$], and the logarithmic $\Sigma_{\rm SFR}$ in units of [M$_{\odot}$\,yr$^{-1}$\,kpc$^{-2}$]. 

The gray and blue background contours indicate the relative number densities of data points used for measuring the H{\sc i} and H$_{2}$ $\Sigma_{\rm GAS}$ of the THINGS galaxies, respectively, spaced in steps of 25\,\%, 50\,\%, 75\,\%, and 100\,\%. $\Sigma_{\rm H2}$ of the THINGS sample galaxies is mainly derived from their central region where CO emission is bright enough to be observed. The dotted lines represent constant gas depletion time ($\tau_{dep}$) of $10^{8}$, $10^{9}$, and $10^{10}$\,yr over which gas is converted to form stars. When estimating $\tau_{dep}$, a factor of 1.4 is multiplied to the sum of H{\sc i} and $\rm H_{2}$ gas surface density to account for helium.

As discussed in \citet{bigiel2008star}, the $\Sigma_{\rm SFR}$ \-- $\Sigma_{\rm H2}$ scaling relation for high mass spiral galaxies is roughly linear in logarithmic scale. This molecular K-S law is well explained by a power law with a slope of {\it N} $\sim 1.00$ in Eq.~(\ref{eq:kslaw}). This implies that stars form where the gas in the galaxies is molecular, and the $\Sigma_{\rm SFR} / \Sigma_{\rm H2}$ (= SFE) in disks appears to be fixed at high gas densities.

On the other hand, in the low gas density regime where most of the hydrogen is in the atomic phase, the $\Sigma_{\rm SFR}$ \-- $\Sigma_{\rm GAS}$ scaling relation does not well follow the trend of the linear extension of the molecular K-S law at low gas densities, having a large range in $\tau_{dep}$. A sharp transition is seen at $\Sigma_{\rm GAS}$ $\sim$\,10 M$_{\odot}$\,pc$^{-2}$ above which hydrogen saturates and is in the molecular phase (\citealt{kennicutt2007star}; \citealt{bigiel2008star}; \citealt{leroy2008star}; \citealt{krumholz2014big}; \citealt{de2019revisiting}).

As for the THINGS sample galaxies, we perform a pixel-by-pixel comparison of the $\Sigma_{\rm SFR}$ and $\Sigma_{\rm HI}$ for NGC 6822 at a typical GMC scale of $\sim$114 pc using the cool-bulk, warm-bulk, cool-non-bulk and warm-non-bulk H{\sc i} gas components decomposed in Section~\ref{sec:sect3}. In this work, we relate the corresponding $\Sigma_{\rm HI}$ of the four kinematic gas components with the $\Sigma_{\rm SFR}$ derived using the GALEX {\it FUV} and WISE {\it MIR} data in Section~\ref{sec:sect2.3}. The FUV-only and hybrid (FUV and MIR) conversions of Eq.(\ref{eq:sfr_both}) and Eq.(\ref{eq:sfr_fuvonly}) trace the star forming activity over a rather long timescale of $\sim$\,100\,Myr. This means that the derived local SFRs do not separate well whether the Type Ia supernovae are on. Therefore, we consider the possibilities of both star formation and stellar feedback when examining the resolved K-S law below. In addition, we exclude the SE part of the galaxy where the H{\sc i} fluxes are significantly affected by the Galactic HI emission as described in Section~\ref{sec:sect2.1}. This is to avoid any artifacts which can be caused by the interpolation process in the resolved K-S law analysis. We indicate the excluded region as a dashed rectangle in the bottom rightmost panel of Fig.~\ref{fig:atlas} (the velocity-dispersion map of the warm-non-bulk gas components). In Fig.~\ref{fig:ks}, we overplot the results which are corrected for the kinematic inclination of 60\,$^\circ$ of NGC 6822 as dots onto the resolved K-S law derived for the THINGS galaxies.

The relations derived for the cool-bulk, warm-bulk, cool-non-bulk, and warm-non-bulk gas components are shown in panels (a), (b), (c), and (d) of Fig.~\ref{fig:ks}, respectively. The dots color-coded from yellow to black in the panels indicate the values measured in different galactocentric radius bins from 0 to 4.5\,kpc in steps of 0.75\,kpc. The error bars shown in each panel of Fig.~\ref{fig:ks} indicate mean values of systematic uncertainties. For the uncertainties of $\Sigma_{\rm GAS}$, we derive the mean value of $bg_{rms}$ of the H{\sc i} line profiles of each kinematic gas component. For the uncertainties of $\Sigma_{\rm SFR}$, we calculate the ones which are propagated from the background rms of the {\it FUV} and {\it MIR} emissions in Eq.~(\ref{eq:sfr_both}). These systematic uncertainties are then converted to the values in units of the $\Sigma_{\rm GAS}$ and $\Sigma_{\rm SFR}$. We also note that an additional uncertainty ($\pm\,15\%$) for the calibration constants in \citet{calzetti2013star} contributes to the systemic uncertainty of $\Sigma_{\rm SFR}$. It corresponds to $\sim$\,0.07 dex in the ordinate of the K-S law.

\begin{figure}
    \centering
    \includegraphics[width=0.9\columnwidth]{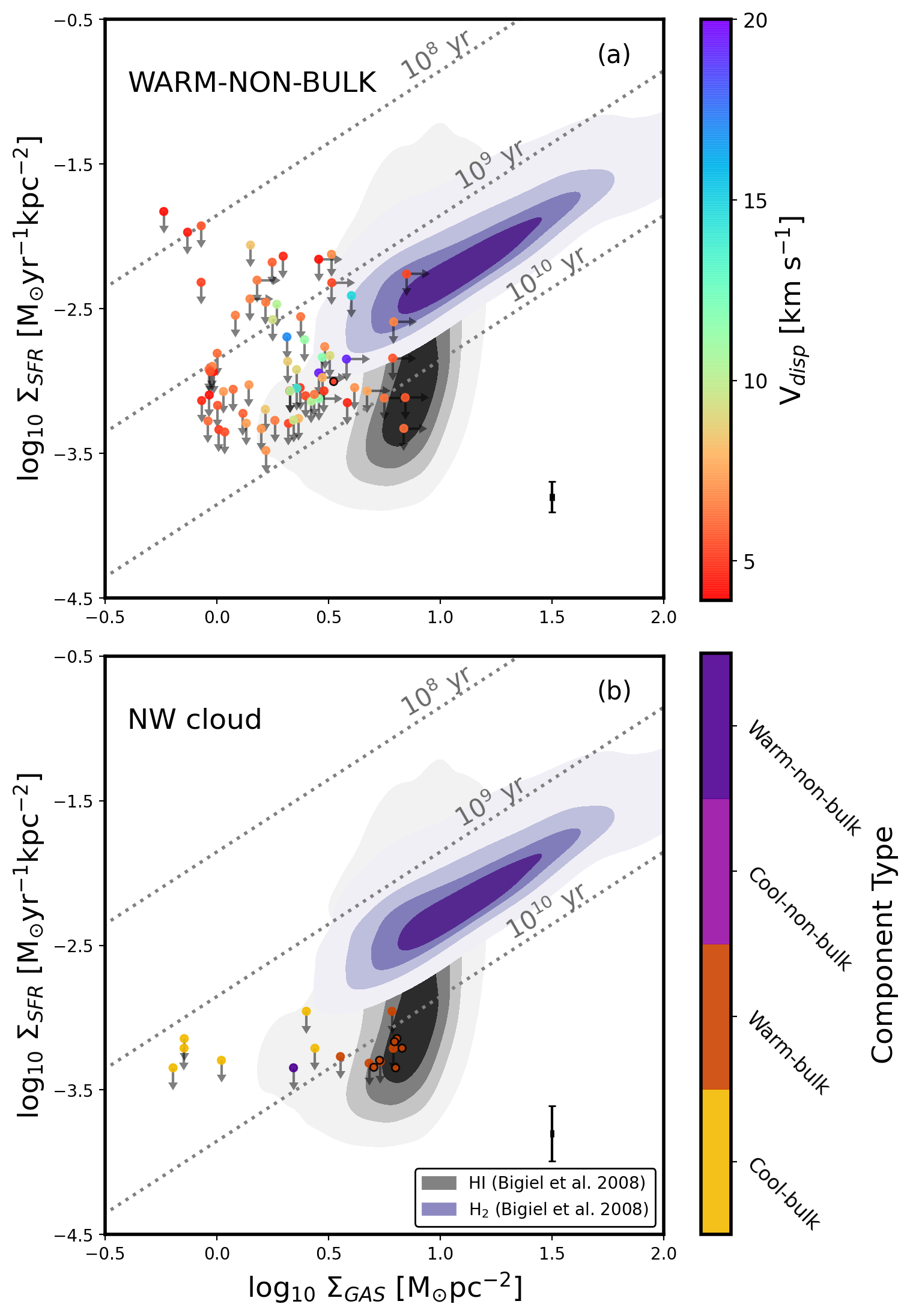}
    \caption{{\bf panel (a)}: The K-S law for the warm-non-bulk gas components varying with its velocity dispersion. {\bf panel (b)}: The K-S law of NW cloud. Yellow and red points present their component type. Other features such as the points with upper limits or black edges, background K-S law, dotted lines and the error bars are the same as in the Fig.~\ref{fig:ks}. The x-axis error for the panel (a) and (b) are 0.007 and 0.006 dex, respectively.}
    \label{fig:add_ks}
\end{figure}

The dots with black outline in Fig.~\ref{fig:ks} and Fig.~\ref{fig:add_ks} represent the gas components whose velocity profiles either best described by a single Gaussian (SGfit) or multiple Gaussians but they are classified in the same kinematic group. Hence, the corresponding $\Sigma_{GAS}$ represent their total gas mass. However, for the others with multiple (two or three) kinematic components that are not in the same kinematic group, the estimated $\Sigma_{SFR}$ should be upper limits. The dots with arrow in Fig.~\ref{fig:ks} and Fig.~\ref{fig:add_ks} apply to this case.

Detection of H$_{2}$ gas components in some regions of NGC 6822 has been reported by other works from SEST\footnote{Swedish-ESO Submillimetre Telescope}, IRAM-30m, and ALMA observations (\citet{israel1997h2}, \citet{gratier2010molecular}, and \citet{schruba2017physical}, respectively). As H$_{2}$ gas components are not considered in our analysis, we note that the $\Sigma_{GAS}$ in the H$_{2}$ detected regions could be underestimated. The corresponding dots are marked by right arrows in Fig.~\ref{fig:ks} and Fig.~\ref{fig:add_ks}.

As shown in the panels (a), (b), (c) and (d) of Fig.~\ref{fig:ks}, the $\Sigma_{\rm SFR}$ values tend to increase with decreasing radius. The derived correlation coefficient for the relationship between $\Sigma_{\rm SFR}$ and galaxy radius is about -0.22, which indicates a weak correlation. This could be mainly because of the higher SFR toward the center of the galaxy, which is also found in other galaxies (\citealt{kennicutt1989star}; \citealt{martin2001star}; \citealt{leroy2008star}). Largely, the $\Sigma_{\rm SFR}$ of the warm-bulk H{\sc i} gas is lower than that of the cool-bulk one. In the context of star formation, this can be due to the higher gas velocity dispersion which might be caused by hydrodynamical processes and gravitational effects in NGC 6822. 
The larger gas turbulence can act as a process that opposes self-gravity, which results in less star formation.
This can be also interpreted by the longer gas depletion time ($\tau_{dep}\!>\!10^{10}$\,yr) for the warm-bulk H{\sc i} gas which is comparable to the Hubble time. The resolved K-S law for the warm-bulk H{\sc i} gas is largely consistent with the one found in previous studies (e.g., \citealt{bigiel2008star}).

Since atomic hydrogen should have passed through a cold phase in the course of forming molecular hydrogen where hydrogen is still in the atomic phase while having a low temperature, part of the cool-bulk H{\sc i} gas may trace this namely, kinematically cold H{\sc i} gas in the vicinity of molecular gas clouds. In this respect, the cool-bulk H{\sc i} gas might act as a gas reservoir which is directly associated with star formation in NGC 6822. Or some of them could be the ones that have been leftover from the star formation process in the galaxy. 

To verify the potential link between the cool-bulk H{\sc i} and $\rm H_{2}$ in the star formation process of NGC 6822, we compare the K-S law for the cool-bulk H{\sc i} gas with the molecular K-S law derived from the THINGS sample galaxies. First, we derive a power-law index {\it N} for the molecular K-S law by fitting Eq.~(\ref{eq:kslaw}) to the central $\Sigma_{H2}$ of the THINGS sample galaxies, i.e., the data points of the blue background contours in Fig.~\ref{fig:ks}. For this, we use an orthogonal distance regression (ODR) method which minimises the sum of perpendicular offsets of data points from a model, taking the errors in both abscissa and ordinate into account (\citealt{brown1990statistical}). Practically, we carry out the model fitting using the {\sc python} module, scipy.odr\footnote{\url{https://docs.scipy.org/doc/scipy/reference/odr.html}}. The derived {\it N} is $\sim$1.01\,$\pm$\,0.15 which is consistent with the one derived in \cite{bigiel2008star}. This is shown as the black dashed line in the cool-bulk K-S law in Fig.~\ref{fig:ks}. The corresponding gas depletion time $\tau_{dep}$ for converting gas into stars is $\sim$\,2.50 $\pm$ 0.38 $\times$\,10$^9$\,yr.

To some extent, the $\Sigma_{\rm SFR}$\--$\Sigma_{\rm HI}$ relation for the cool-bulk H{\sc i} gas (panel (a) of Fig.~\ref{fig:ks}) is weakly but better consistent with the linear extension of the molecular K-S law at low gas densities compared to the other kinematic components (i.e., warm-bulk, cool-non-bulk, warm-non-bulk). The estimated mean gas depletion time and rms of the components are denoted on the bottom-left corner of the panels in Fig.~\ref{fig:ks}. We find a mean value of $\tau_{dep}$\,$\sim$\,2.06 $\pm$ 1.81 $\times$\,10$^9$\,yr for the cool-bulk H{\sc i} gas which is comparable to the one for the molecular gas. This trend becomes more evident for the cool-bulk H{\sc i} gas in the galaxy with increasing radius despite the lack of data points. In the outer region of the galaxy where $\Sigma_{\rm HI}$ is low ($\Sigma_{\rm HI}<$\,10\,M$_{\odot}$\,pc$^{-2}$), the kinematically cold H{\sc i} gas can be an useful tracer for star formation which is predicted from the molecular K-S law.

The warm-non-bulk H{\sc i} gas has a wide range of $\tau_{dep}$ in the $\Sigma_{\rm SFR}$\--$\Sigma_{\rm HI}$ relation, and some of them have even smaller $\tau_{dep}$ than that of the cool-bulk H{\sc i} gas. This could be a net result both from positive and negative effects of either hydrodynamical processes or gravitational interactions on star formation in NGC 6822. As defined in Section~\ref{sec:sect4}, the non-bulk H{\sc i} gas deviates from the global kinematics of NGC 6822 which is derived from a 2D tilted-ring analysis of the H{\sc i} {\sc moment1} in \cite{weldrake2003high}. As discussed in Section~\ref{sec:sect5}, stellar feedback associated with star formation and/or gravitational interactions in a galaxy can impact ambient ISM in a way that ionises H{\sc i}, gives rise to radiation pressure and/or injects kinetic energy (\citealt{hopkins2018how}). Consequently, these hydrodynamic and gravitational effects can lead to either quenching or triggering of star formation in galaxies (\citealt{man2018star}; \citealt{inutsuka2015formation}). This could be the case of the $\Sigma_{\rm SFR}$\--$\Sigma_{\rm HI}$ relation for the warm-non-bulk H{\sc i} gas whose kinematics might be affected by such radiative and/or mechanical feedback in NGC 6822. For these deviating broad H{\sc i} gas clouds, $\Sigma_{\rm SFR}$ is not well correlated with $\Sigma_{\rm HI}$ regardless of gas velocity dispersion (See the panel (a) of Fig.~\ref{fig:add_ks}). Within these gas clouds, star formation might be regulated by the combined effect of local processes like turbulence, stellar feedback and metallicity (e.g., \citealt{krumholz2005formation}) or even galactic-scale processes like cloud-cloud interactions (e.g., \citealt{silk1997feedback}; \citealt{gong2017promising}; \citealt{fukui2021cloud}). 

In addition, we note that the classification of the non-bulk H{\sc i} gas made in Section~\ref{sec:sect4} is dependant on the two main factors, the model reference velocity field for the global kinematics of NGC 6822 and the velocity limit used for distinguishing bulk and non-bulk H{\sc i} gas motions. As we use $1\times$ the SGfit velocity dispersion for the velocity limit which is likely to be conservative, some of bulk H{\sc i} gas might be misclassified as the non-bulk one, and vice versa. This might result in the spread of the $\tau_{dep}$ in the $\Sigma_{\rm SFR}$\--$\Sigma_{\rm HI}$ relation for the non-bulk H{\sc i} gas in Fig.~\ref{fig:ks}.

Lastly, the $\Sigma_{\rm SFR}$\--$\Sigma_{\rm HI}$ relation for the NW H{\sc i} cloud at a galactocentric distance of R\,$\sim$\,4\,kpc is represented in panel (b) of Fig.~\ref{fig:add_ks}. The points are colorized by the type of the component as yellow, orange, magenta, purple for cool-bulk, warm-bulk, cool-non-bulk, and warm-non-bulk components, respectively. As discussed earlier, the H{\sc i} gas in and around the region is likely to be affected by recent star formation activities and/or gravitational interaction between the cloud and NGC 6822. However, the $\Sigma_{\rm SFR}$\--$\Sigma_{\rm HI}$ relations for the cool-bulk, warm-bulk, cool-non-bulk and warm-non-bulk H{\sc i} gas components in the vicinity of the NW H{\sc i} cloud region are not much different from their respective relations that are derived for other regions except that only a few data points are available for the non-bulk one. 

\vspace{0.1mm}

\section{Summary}
\label{sec:sect7}
In this paper, we investigate the gas kinematics and star formation activities of a nearby dwarf galaxy, NGC 6822 in the local group at a distance of $\sim$\,490 kpc using the multi-wavelength data of the galaxy including the ATCA H{\sc i} 21cm, GALEX {\it FUV}, and WISE-W4 {\it MIR} data. The proximity and the extended H{\sc i} gas disk of the galaxy located in a relatively isolated environment make it an ideal object to examine the potential effect of internal hydrodynamic processes and gravitational interactions on star formation as well as on the kinematics of ambient ISM.

We first decompose all the line profiles of the ATCA H{\sc i} data cube with optimal numbers of Gaussian components using a tool, the so-called {\sc baygaud} (\citealt{oh2019robust}; \citealt{oh2022kinematic}) which is based on a Bayesian analysis technique. After convolving the raw data cube using a 2D Gaussian kernel  to make its spatial resolution 48.0\arcsec $\times$ 48.0\arcsec, we re-sample the cube with a pixel scale of 48.0\arcsec\,pixel$^{-1}$ which corresponds to a physical scale of $\sim$114 pc. This is for making individual H{\sc i} line profiles of the cube independent with each other in the course of the profile decomposition. As found in \citet{de2006star} and this study, significant fraction of the line profiles is best described by Gaussian models with a sum of up to three Gaussians, particularly in the inner region of the H{\sc i} disk, in the vicinity of the supergiant shell, and around the NW cloud region.

Of the decomposed Gaussian components, we classify those that move within a velocity limit against the global kinematics of NGC 6822 as bulk H{\sc i} gas components. The rest of the Gaussian components deviating from the global kinematics by more than a velocity limit is classified as non-bulk H{\sc i} gas components. As a conservative choice for the velocity limit, we use one times SGfit velocity dispersion. For the global kinematics of NGC 6822, we construct a model velocity field from the 2D tilted-ring analysis of the {\sc moment1} map which is presented in \citet{weldrake2003high}. Subsequently, we further classify the bulk and non-bulk H{\sc i} gas components as narrow and broad ones whose velocity dispersions are less or greater than 4 \kms\ which is determined from the velocity dispersion histogram analysis using the bulk Gaussian components. We call these as cool-bulk, warm-bulk, cool-non-bulk and warm-non-bulk H{\sc i} gas components, respectively.

The warm-bulk H{\sc i} gas with a mean velocity dispersion of 8.9\,\kms\ is distributed throughout the disk, showing a somewhat regular rotation pattern in the velocity field except for the supergiant shell region inside which a giant H{\sc i} hole resides. Some of cool-bulk H{\sc i} gas components appear to be located in the vicinity of the NW cloud region and the supergiant shell as well as along the minor axis of the disk as for the warm-non-bulk H{\sc i} gas. As traced by the {\it FUV} and {\it MIR} emissions, some of the cool-bulk H{\sc i} gas are likely to be associated with star formation in the galaxy which act as the gas reservoir of star formation fuel. On the other hand, part of the turbulent warm-bulk and warm-non-bulk H{\sc i} gas could be caused by stellar feedback in the galaxy.

We derive the total H{\sc i} gas mass of $1.29\,\times\,10^{8}$\,M$_{\odot}$ in NGC 6822 using the optimally decomposed Gaussian components of the H{\sc i} data cube. This is $\sim$\,7\% higher than the H{\sc i} mass derived using the {\sc moment0} map. Compared to the conventional analysis such as the {\sc moment0} and SGfit fitting method, via the profile decomposition, we were able to retrieve more mass of the H{\sc i} gaseous components in low ($<$ 4.0\,\kms) velocity dispersion bins. Our profile decomposition analysis was able to de-blend asymmetric non-Gaussian velocity profiles of the cube, and extracts more Gaussian components with lower velocity dispersions.

We perform a pixel-by-pixel analysis of the multi-wavelength data set (i.e., the H{\sc i}, {\it FUV} and {\it MIR} data) which relates the surface densities of the kinematically decomposed gaseous components ($\Sigma_{\rm HI}$) to their corresponding SFR surface densities ($\Sigma_{\rm SFR}$). We derive the $\Sigma_{\rm SFR}$ using the GALEX {\it FUV} and WISE-W4 {\it MIR} images of the galaxy which trace the unobscured and obscured SFRs, respectively. For this, we apply the conversion factors which are given in \citet{calzetti2007calibration}, \citet{liu2011super} and \citet{calzetti2013star} to the {\it FUV} and {\it MIR} images whose pixel scales are degraded to match the one of the smoothed H{\sc i} data cube (i.e., 48.0\arcsec\,pixel$^{-1}$). The total SFR of NGC 6822 estimated using the re-sampled {\it FUV} and {\it MIR} images is $\sim$\,0.015\,M$_{\odot}$\,yr$^{-1}$ which is consistent with the ones presented in other studies (\citealt{hunter2010galex}; \citealt{efremova2011recent}).

We derive the resolved K-S law for the cool-bulk, warm-bulk, cool-non-bulk and warm-non-bulk H{\sc i} gas components, and investigate how they change with respect to their radial distance from the kinematic center of NGC 6822: 1) In general, as found in other studies, the closer to the galactic center, the higher the SFR surface densities, 2) The K-S relation behaviour for the warm-bulk H{\sc i} gas is consistent with those found in previous pixel-pixel analysis of nearby galaxies, showing a sharp saturation of H{\sc i} gas surface densities at around $\Sigma_{\rm HI}<$\,10\,M$_{\odot}$\,pc$^{-2}$ below which the relation has a wide range in the gas depletion time $\tau_{dep}$ at a given gas surface density, 3) The K-S relation for the cool-bulk H{\sc i} gas is weakly consistent with the linear extension of the molecular K-S law at low gas surface densities, showing a narrow range of $\tau_{dep}$, 4) Little correlation between the $\Sigma_{\rm SFR}$\--$\Sigma_{\rm HI}$ of the warm-non-bulk H{\sc i} gas is found, showing a wide range in $\tau_{dep}$, which may be due to the net effect of positive and negative stellar feedback in the galaxy, and 5) The K-S relation for the cool-non-bulk H{\sc i} gas is unclear given the lack of available data points although it is likely to show a similar trend as for the cool-bulk H{\sc i} gas; Some of the decomposed H{\sc i} Gaussian components might have been mis-classified as non-bulk H{\sc i} gas depending on the assumed global kinematics of NGC 6822 and the velocity limit, which contributes to the scatter in the K-S relation. To sum up, warm-bulk H{\sc i} components are likely to include not only the H{\sc i} gas which acts as gas reservoir for star formation but also the ones being affected by star formation activities in the galaxy. On the other hand, cool-bulk ones are likely to be closely associated with molecular gases. The non-bulk ones appear to trace star formation activities in the galaxy.

\section*{Acknowledgements}
We would like to thank the anonymous reviewer for her/his careful reading of the manuscript and many insightful and constructive comments and suggestions. 

SHOH acknowledges a support from the National Research Foundation of Korea (NRF) grant funded by the Korea government (Ministry of Science and ICT: MSIT) (No. NRF-2020R1A2C1008706). This work has received funding from the European Research Council (ERC) under the European Union’s Horizon 2020 research and innovation programme (grant agreement No. 882793 ``MeerGas"). Parts of this paper are based on the author's MSc. thesis at Sejong University.
\software{astropy \citet{astropy}, seaborn \citet{seaborn}, CASA \citet{casa}, GIPSY \citet{gipsy}, Photutils \citet{photutils}, Scipy \citet{virtanen2020scipy}}

\bibliography{sample631}{}
\bibliographystyle{aasjournal}

\end{document}